\documentclass[12pt]{article}
\pdfoutput=1
\usepackage{graphics, color}
\usepackage{graphicx}
\usepackage{amsmath}
\usepackage{amssymb}
\usepackage{amsfonts}

\newcommand{\sect}[1]{\section{#1}\setcounter{equation}{0}}

\def\gsim{\, \rlap{$>$}{\lower 1.1ex\hbox{$\sim$}}\,}
\def\lsim{\, \rlap{$<$}{\lower 1.1ex\hbox{$\sim$}}\,}

\newcommand{\be}{\begin{equation}}
\newcommand{\ee}{\end{equation}}
\newcommand{\bea}{\begin{eqnarray}}
\newcommand{\eea}{\end{eqnarray}}

\newcommand{\rmx}{\rm}
\newcommand{\rs}{r_{\rmx s}}
\newcommand{\lp}{l_{\rmx P}}
\newcommand{\la}{l_{\rm AdS}}

\begin{document}


\begin{titlepage}
\bigskip
\bigskip\bigskip\bigskip
\centerline{\Large The Black Hole Information Problem}
\bigskip\bigskip\bigskip
\bigskip\bigskip\bigskip

 \centerline{ {\bf Joseph Polchinski}\footnote{\tt joep@kitp.ucsb.edu}}
\medskip
\centerline{\em Kavli Institute for Theoretical Physics}
\centerline{\em University of California}
\centerline{\em Santa Barbara, CA 93106-4030}\bigskip
\bigskip
\bigskip\bigskip


\begin{abstract}
The black hole information problem has been a challenge since Hawking's original 1975 paper.  It led to the discovery of AdS/CFT, which gave a partial resolution of the paradox.  However, recent developments, in particular the firewall puzzle, show that there is much that we do not understand.  I review the black hole, Hawking radiation, and the Page curve, and the classic form of the paradox.  I discuss AdS/CFT as a partial resolution.  I then discuss black hole complementarity and its limitations, leading to many proposals for different kinds of `drama.'  I conclude with some recent ideas.

Presented at the 2014-15 Jerusalem Winter School and the 2015 TASI.
\end{abstract}
\end{titlepage}
\baselineskip = 17pt
\setcounter{footnote}{0}

\tableofcontents

\sect{Introduction: Quantizing gravity}

The big question that underlies these lectures is `What is the theory of quantum gravity?'  String theory gives a solution to the renormalizability problem, and it does many other good things, but it is only a perturbative expansion in powers of the coupling.  We know that quantum theories exhibit many fascinating and important phenomena that are not visible in perturbation theory.  In the Standard Model, these include quark confinement, chiral symmetry breaking, and electroweak baryon and lepton number violation, and in quantum gravity things should be even more interesting.  So we need to know, what is the exact theory for which string theory is the perturbation expansion? 

The same sort of question was faced in quantum field theory, and the answer was basically given by Wilson as the path integral defined via the renormalization group~\cite{Wilson:1993dy}.   (Though there are still indications, in particular the inability to explain dualities with the path integral approach, that something more is needed.)  The most direct analog for string theory would be string field theory, but in spite of some nice structures such as~\cite{Witten:1985cc}, it does not seem to have answered the question.

We now have a partial answer by way of dualities.  The BFSS matrix model~\cite{Banks:1996vh} and AdS/CFT duality~\cite{Maldacena:1997re,GKP,W} relate string theory in certain backgrounds to quantum mechanical systems and quantum field theories.  These we know how to define, and in principle they contain the whole of  flat spacetime string perturbation theory, and its exact completion.  Joao gave a nice review of AdS/CFT at the 2015 TASI.  The most obvious limitation of these constructions is that they are restricted to string theory in spaces with special boundary conditions, and cannot describe cosmological spacetimes or realistic compactifications.  These dualities have taught us much about quantum gravity, but we still do not know how to generalize these lessons, and it seems likely that we are still missing important concepts.

So we have the question, what is the full quantum theory?  There is the dynamical question: what are the fundamental variables?  But there are a deeper questions.  What is the right framework for the theory?  Do we begin with a Hilbert space?  Are observations connected with the wavefunction in the usual way?  
For special spacetimes, like anti-de Sitter, we have good answers to some of these questions, but it is not clear how these carry over to cosmology.  And now it seems that we have to confront them already  to understand the black hole interior.

Thought experiments with black holes have been essential in getting us to our current understanding.  Considerations of black hole entropy  led 't Hooft and Susskind to infer the holographic principle~\cite{'tHooft:1993gx,Susskind:1994vu}.  The further study of black hole entropy~\cite{Strominger:1996sh} and of information loss~\cite{Hawking:1976ra} then led to an understanding of the relation between black branes and D-branes and so to the BFSS theory and AdS/CFT duality.  

So do we have more to learn from black holes?  As I will discuss, BFSS and AdS/CFT imply that information is not lost.  But there is still the open question, where was Hawking's original argument wrong?  As we will see, Hawking's conclusion seems inevitable if gravity as an effective field theory is valid where one would normally expect an effective theory to be.  Building on the work of others, especially Samir Mathur, AMPS has argued that if information is not lost then effective field theory must break down in a rather dramatic way at the black hole horizon~\cite{Almheiri:2012rt}.  
Attempts to evade this firewall argument, or to confirm it, have lead in many directions and generated many new ideas.  The AMPS argument also exposes unexpected limitations of AdS/CFT duality, in its inability to tell us what is happening in the black hole interior.

There is a striking convergence of interest in quantum entanglement, from the black hole information problem, from Ryu and Takayanagi's holographic entanglement entropy~\cite{Ryu:2006bv} (where it relates spacetime geometry to the dual field theory), from condensed matter physics (where it classifies exotic phases) and from quantum information theory (where it is a resource).  It is not clear where this is going, but there is a sense that we will learn something important from it.\footnote{I was last at TASI in 2010, where applications of AdS/CFT, including condensed matter, were a main focus, and remarkably I cannot recall a single mention of entanglement then.}

In \S2-3 we review standard background on black holes and Hawking radiation.  In \S4 we review the information loss problem, focusing on the Page curve, and we present the three classing alternatives.  In \S5 we review AdS/CFT and its black holes, and argue that it excludes information loss and remnants.   We also  take an excursion to ask the question, can we tell if a black hole is in a pure state, and we introduce the swap test.
In \S6 we introduce black hole complementarity as a proposed resolution to the information long problem, and explain why it falls.  In \S7 we consider some alternatives: modification of the geometry at the Schwarzschild distance, modification of the dynamics and geometry near the horizon due to fuzz, fire, or strings, or modification of quantum mechanics through the final state or limits on quantum computation.  In \S8 we expand on the question, how to we see the bulk physics from the CFT, in particular behind the horizon.  The leads to two more possible modifications, state-dependence and ER=EPR.
In \S9 we discuss some recent related ideas, and conclude.

Another recent review of this subject is~\cite{Harlow:2014yka}.  Those lectures are perhaps more modern, emphasizing the connections with quantum information theory, while here I have tried to collect many traditional arguments that are scattered in many places.  Older reviews include~\cite{Page:1993up,Mathur:2009hf}.

\sect{Black holes}

Begin with the Schwarzschild metric\footnote{I will not be looking at rotating or charged black holes explicitly.  The near-horizon geometry is the same, away from the extremal limit.}
\be
ds^2 = - \left( 1 - \frac{r_{\rmx s} }{r} \right) dt^2 +  \left( 1 - \frac{r_{\rmx s} }{r} \right)^{-1} dr^2 + r^2 (d\theta^2 + \sin^2\theta d\phi^2)\,,  \label{schw}
\ee
with Schwarzschild radius $r_{\rmx s}  = 2GM$.  I will work in units $\hbar = c = k_{\rmx B}= 1$, with Newton's constant $G = l^2_{\rm P}= m^{-2}_{\rm P}$ explicit (${\rm P} =  {\rm Planck}$). 

Due to the gravitational redshift, much of the important dynamics occurs close to the horizon.  We  thus expand the metric in this region, $r =  r_{\rmx s} + \delta$.
The small-$\delta$ behavior is
\be
ds^2 \cong - \frac{\delta}{ r_{\rmx s}} dt^2 + \frac{ r_{\rmx s}}{\delta} d\delta^2 +  r_{\rmx s}^2 (d\theta^2 + \sin^2\theta d\phi^2)\,.
\ee
Defining $\delta = \rho^2/4  r_{\rmx s}$ this becomes
\be
ds^2 \cong - \frac{\rho^2}{4 r_{\rmx s}^2} dt^2 +  d\rho^2 +  r_{\rmx s}^2 (d\theta^2 + \sin^2\theta d\phi^2)\,.
\ee

To make this look more familiar, and because we will need it later, let us go to Euclidean time, $t_{\rmx E} = it$:
\be
ds^2 \cong  \frac{\rho^2}{4 r_{\rmx s}^2} dt_{\rmx E}^2 +  d\rho^2 +  r_{\rmx s}^2 (d\theta^2 + \sin^2\theta d\phi^2)\,.
\ee
We recognize the first two terms as two-dimensional flat space.  Defining 
\be
X = \rho \cos (t_{\rmx E}/2r_{\rmx s}), \quad Y = \rho \sin (t_{\rmx E}/2r_{\rmx s})\,, \label{xy}
\ee
the metric is
\be
ds^2 \cong  dX^2 + dY^2 +  r_{\rmx s}^2 (d\theta^2 + \sin^2\theta d\phi^2)\,.
\ee
Similarly for the Lorentian case define 
\be
X = \rho \cosh (t/2r_{\rmx s}), \quad T = \rho \sinh (t/2r_{\rmx s})\,,
\ee
 and then
\bea
ds^2 &\cong&  -dT^2 + dX^2 +  r_{\rmx s}^2 (d\theta^2 + \sin^2\theta d\phi^2)  \nonumber\\
&=& -dU dV +  r_{\rmx s}^2 (d\theta^2 + \sin^2\theta d\phi^2)  
\,, \label{NH}
\eea
where 
\be
U = T-X = - \rho e^{-t/2r_{\rmx s}}, \quad V= T+X = \rho e^{t/2r_{\rmx s}} \label{uv}
\,.
\ee

The Schwarzschild coordinates $r,t$ cover only the region $X > |T|$ (so $U<0, V>0$), known as quadrant 1.  This geometry can be smoothly to the four regions
\bea
{\rm quadrant }\ 1:\ \ \  \ \  & U<0,\ V>0, \ \ \ {\rm Schwarzschild}  \nonumber\\
{\rm quadrant }\ 2:\ \ \  \ \  & U>0,\ V>0,  \ \ \ {\rm future\ interior} \nonumber\\
{\rm quadrant }\ 3:\ \ \  \ \  & U>0,\ V<0,  \ \ \ {\rm Schwarzschild'} \nonumber\\
{\rm quadrant }\ 4:\ \ \  \ \  & U<0,\ V<0,  \ \ \ {\rm past\ interior}\,.  \label{quad}
\eea

Going back to the full metric, the corresponding extension is\footnote{There is a finite rescaling here, $(U,V)_{(\ref{krus})} =  \frac12 e^{1/2} (U,V)_{(\ref{NH})}$. }
\bea
ds^2 &=& -\frac{4 r_{\rmx s}}{r} e^{-r/r_{\rmx s}} dU dV + r^2 (d\theta^2 + \sin^2\theta d\phi^2) \,, 
\nonumber\\
UV &=& \rs ( {r_{\rmx s}  - {r} }) e^{r/\rs} \,,\quad \frac{U}{V}= -e^{-t/r_{\rmx s}}. \label{krus}
\eea
The limit $r \to 0$ is a future singularity in quadrant 2 and a past singularity in quadrant 4.
The full metric describes two black holes, regions I and III, joined by an Einstein-Rosen bridge.
The future horizon is 

We see from the coordinates~(\ref{xy}) that in order for the Euclidean metric to be smooth, $t_{\rm E}$ must be periodic with period $4\pi r_{\rmx s}$.  A path integral with periodic Euclidean time $t_{\rm E} \sim t_{\rm E} + \beta$ generates the thermal partition function Tr$\,e^{-\beta H}$.
The  path integral for quantum fields in the Euclidean black hole geometry thus describes a gas at temperature $T_{\rmx H} = 1/4\pi r_{\rmx s}$ in equilibrium with the black hole.  
{\it The black hole has a temperature, and it must be able to emit as much as it absorbs.}  Moreover, by thermodynamic relations, the corresponding entropy is
\be
dS = \frac{dM}{T_{\rmx H}} = -\frac{dT_{\rmx H}}{8\pi G  T_{\rmx H}^3} \ \Rightarrow\ S = S_{\rmx BH} \equiv \frac{\pi r_{\rmx s}^2}{G} = \frac{A}{4G} \,,
\ee
where $A$ is the horizon area.

The Euclidean black hole continues to a two-sided Lorentzian black hole.
You have heard much about the two-sided story from Juan and Mark.  However, the original information problem arises in the collapse of a single black hole, and we will often focus on this case.
For these, there will be some infalling matter following a timelike or lightlike trajectory from region I to region II and to the future singularity, and everything to the left of this is replaced by the smooth interior geometry of the infalling matter.  Note that time translations stretch $V$ and scrunch $U$ (or the reverse), so from the point of view of a late-time observer the infalling matter is very close to the null line $V=0$.

In statistical mechanics, the exponential of the entropy is a count of the states available to a system.  What are these states here?  Bekenstein provided an approximate count of them as follows~\cite{Bekenstein:1973ur}.  Imagine throwing quanta into a black hole.  In order to fit, their size must be no larger than $\rs$ and so their energies are at least $1/\rs$.  For a black hole of mass $M$, the total number of quanta grows no more rapidly than 
\be
dN \sim \rs dM \sim \rs d\rs/G  \,.
\ee
With $O(1)$ unit of entropy per quantum, this gives $S \sim A/G$ as found above.  So according to this picture, we can identify the states with those of infalling matter inside the black hole.  Indeed, we can consider a nice spacelike slice that interpolates from a fixed $t$ slice on the outside to a fixed $r < \rs$ slice in the interior.  As the black hole gets older, this slice gets longer on the inside, and all the infalling matter is present on it.  So the microstates of the black hole seem to correspond to the states of the fields on the nice slice.  It is good to keep this picture in mind, but it will run us into problems.\footnote{The main problem will arise in the context of quantum mechanics, but already classically it is a bit problematic.  By throwing in the quanta very slowly, we can encode information in the timing, and so encode many bits per quantum.}

\sect{Hawking radiation}

Let us understand better how the black hole can emit radiation.  Consider an inertial observer falling through the (future) horizon.  This takes a finite amount of time $\tau$ in their own frame, but an asymptotic observer never sees them on the horizon for an infinite time $t$, so there is a highly nonlinear relation between their times (more precisely, between the times that they encounter a given outgoing null ray).  In terms of the coordinates~(\ref{uv}), $V$ is approximately constant while $U$ is going through zero linearly in the proper time $\tau$ of the infalling observer.  This implies that
\be
d\tau \propto e^{-t/\rs} dt \,.
\ee

The coordinate $\tau$ (linear in $U, V$)  smoothly crosses the ingoing  horizon 
\be
U = 0 \,,\quad V = {\rm const} \,.  
\ee
The coordinate $t$ stops at the horizon.   An observer using the $\tau$ coordinate can cross the horizon freely.  An observer using the $t$ coordinate will interpret space as ending at the horizon.   
The relation between these two coordinates essentially generates the whole story, from the Hawking radiation to the paradoxes that it produces.

 If the infalling observer expands a quantum field in modes of given $\tau$-frequency $\nu$, and the outside observer expands the fields in modes of given $t$-frequency $\omega$,  these are not the same expansion, and in particular positive and negative frequencies get mixed.  Also, if we consider a mode of fixed $\omega \sim \rs^{-1}$, then the earlier the infalling observer meets the mode, the higher its typical frequency $\nu$ will be in that frame.    In other words, the mode blue-shifts as we follow it backward, or red-shifts as we follow it forward.

Consider foliating the near-horizon geometry with smooth slices, for example taking $U+V$ as the time coordinate.  
In this foliation, the geometry is changing adiabatically on a time scale $\rs^{-1}$, but the modes we are discussing have much higher frequency.  The geometry is changing slowly compared to $\nu$, so by the adiabatic principle this mode must be in its ground state, to high accuracy $e^{-O(\nu\rs)}$, in the modes of the infalling observer.  {If the Hamiltonian for a quantum system is changing at a rate slow compared to the spacing between levels, then the probability for the system to become excited is exponentially small.} 

Now let us flesh this out a bit.  For extreme simplicity we ignore the angular directions and treat this as a 1+1 dimensional system with a massless scalar $\phi$.  The metric is
\bea
ds^2 &=& - \left( 1 - \frac{r_{\rmx s} }{r} \right) dt^2 +\left( 1 - \frac{r_{\rmx s} }{r} \right)^{-1} dr^2  \,, \nonumber\\
 &=& - \left( 1 - \frac{r_{\rmx s} }{r} \right) du\, dv  \,, \nonumber\\
&=& -\frac{4 r_{\rmx s}^2}{r} e^{-r/r_{\rmx s}} dU dV \,.
\eea
In the second line we have introduced 
\be
u = t-r_* = -2\rs \ln (-U/\rs)\,,\quad v= t+r_* = 2\rs\ln (V/\rs)\,,
\ee
where $r_* = r + \rs \ln(r-\rs)$.  The coordinates $u,v$ are conformally related to $U,V$ but are defined only in quadrant I; they are the null coordinates for the asymptotic observer.  The Klein-Gordon equation is simply 
\be
\partial_u \partial_v \phi = \partial_U \partial_V \phi = 0  \,,
\ee
giving right $+$ left-moving modes in either coordinate system.
We will only need the right-moving part, which we expand in modes,
\bea
 \phi_{R} &=& \int_{0}^\infty \frac{d\nu}{2\pi(2\nu)^{1/2}}\, \left( a_\nu e^{-i\nu U} +a_\nu^\dagger e^{i\nu U} \right) \nonumber\\
&=& \int_{0}^\infty \frac{d\omega}{2\pi(2\omega)^{1/2}}\, \left( b_\omega e^{-i\omega u} +b_\omega^\dagger e^{i\omega u} \right)   \,.
\eea
The nonzero canonical commutators are
\be
[ a_\nu, a_{\nu'} ] = 2\pi \delta(\nu-\nu')\,,\quad [ b_\omega, b_{\omega'} ] = 2\pi \delta(\omega-\omega') \,.
\ee

The coordinate $U$ is smooth across the horizon, so the $a_\nu$ are good modes for the infalling observer.  For the asymptotic observer, one wants to use the modes $b_\omega$ of definite frequency with respect to the time translation symmetry (we are considering times well after the collapse,  ignoring transient effects).  From the mode expansions, the relation between these is 
\be
b_\omega = \int_{0}^\infty \frac{d\nu}{2\pi}\left( \alpha_{\omega\nu} a_\nu +  \beta_{\omega\nu} a_\nu^\dagger \right)\,, \label{atob}
\ee
where
\bea
\alpha_{\omega\nu} &=& 2\rs (\omega/\nu)^{1/2} (2 \rs\nu)^{2i  \rs \omega} e^{\pi  \rs \omega} \Gamma(- 2i  \rs \omega) \,, \nonumber\\
\beta_{\omega\nu} &=& 2\rs (\omega/\nu)^{1/2}  (2 \rs\nu)^{2i  \rs \omega} e^{-\pi  \rs \omega} \Gamma(- 2i  \rs \omega) \,.
\eea

By the adiabatic principle, the horizon-crossing modes $a_\nu$ approach the 
 black hole vacuum state, satisfies $a_\nu | \psi \rangle = 0$.  The eternal modes then behave as
\bea
\langle \psi | b_\omega^\dagger b_{\omega'} | \psi \rangle &=& 2(\omega\omega')^{1/2}\int_{0}^\infty \frac{d\nu}{2\pi(2\nu)^{1/2}}\frac{d\nu'}{2\pi(2\nu')^{1/2}}
\,\beta_{\omega\nu}^*\beta_{\omega'\nu'}  \langle \psi | a_\nu a_{\nu'}^\dagger | \psi \rangle \nonumber\\
&=& 2(\omega\omega')^{1/2}\int_{0}^\infty \frac{d\nu}{4\pi \nu}
\,\beta_{\omega\nu}^*\beta_{\omega'\nu}  \nonumber\\
&=& \frac{2\pi \delta(\omega-\omega')}{e^{4\pi  \rs \omega} - 1} \nonumber\\
&=& \frac{2\pi \delta(\omega-\omega')}{e^{\omega/T_{\rmx H}} - 1}  \,. \label{hflux}
\eea
This is a blackbody spectrum of the expected temperature.  

The original derivation of Hawking~\cite{Hawking:1974sw} involved integrating the modes through the infalling body.  This is unphysical: the interaction between the modes and the matter is hyper-Planckian due to the blueshift effect.  But Hawking understood that it was only the horizon structure that mattered, essentially what we have called the adiabatic principle.  The importance of the adiabatic principle was emphasized by Jacobson~\cite{Jacobson:2003vx}; it was also employed in~\cite{Mukhanov:2007zz}.
The adiabatic principle is also used in cosmology, to initialize the state of the quantum fluctuations.  At both the black hole horizon and in the expanding universe, modes are blueshifting as we move back in time, and if one follows them far enough then they begin far in the UV where we do not know the physics but we still assume the adiabatic principle to hold.

The coordinates $u,v$, and so the modes $b_\omega$, are defined only in quadrant \mbox{I}.  This was enough to derive the Hawking flux~(\ref{hflux}),
but to discuss the interior  we need to introduce the corresponding modes in quadrant \mbox{II}.\footnote{For a more complete discussion along the same lines, see~Ref.~\cite{Giddings:1992ff}.}
  Denote these interior modes by $\tilde b_\omega$.   Since the $a_\nu$ cross the horizon, the inverse of the mode relation~(\ref{atob}) must involve the tilded modes as well,
\be
a_\nu = \int_{0}^\infty \frac{d\omega}{2\pi}\left( \alpha^*_{\omega\nu} b_{\omega} - \beta^*_{\omega\nu} b_{\omega}^\dagger 
+  \tilde\alpha^*_{\omega\nu} \tilde b_{\omega} - \tilde\beta^*_{\omega\nu} \tilde b_{\omega}^\dagger \right)\,. \label{btoa}
\ee
Using this we can write the $a$-vacuum (the state annihilated by the $a_\nu$) in terms of the $b$- and $\tilde b$-vacua.  

We can deduce the final state rather simply:
\be
|0\rangle_a = {\cal N} \exp \left( \int _0^\infty  \frac{d\omega}{2\pi} \, e^{- \omega/2T_{\rmx H}}  b_\omega^\dagger \tilde b_\omega^\dagger \right),
\label{bogo}
\ee
where ${\cal N}$ is a normalization factor.\footnote{The difference between the simple vacuum $|0\rangle_a$ and the generic black hole state $|\psi \rangle$ in~(\ref{hflux}) is that the latter describes additional degrees of freedom as well, such as the ingoing modes.}
  The squeezed form follows from the relation~(\ref{btoa}).  Much of the rest is determine by the requirement that this respect the time-translation symmetry of the black hole (again, the whole process is happening long after the formation of the black hole, so that the geometry is effectively static).  Now, the mode operator $b_\omega^\dagger$ raises the energy,
\be
[H, b_\omega^\dagger] = \omega b_\omega^\dagger \,.
\ee
If $\tilde b_\omega^\dagger$ satisfies the same relation then  $b_\omega^\dagger \tilde b_\omega^\dagger$ would not commute with $H$ and this could not work; in fact, no nontrivial form would be allowed.  But in fact if one traces through one finds that $\tilde b_\omega^\dagger$ must be defined so as to lower the energy,
\be
[H, \tilde b_\omega^\dagger] = -\omega b_\omega^\dagger \,, \label{negen}
\ee
and then the form~(\ref{bogo}) is fully fixed, with the coefficient in the exponent determined by the flux~(\ref{hflux}).  The form~(\ref{bogo}) means that the $b$ and $\tilde b$ modes are entangled with each other, a fact that will play a central role as we go along.

The negative energy~(\ref{negen}) might seem odd.  The point is that what we are calling energy is the conserved charge associated with the Killing vector that looks like time translation outside the black hole.  This Killing vector changes signature at the horizon and so is actually a momentum for the interior modes, and so either sign is allowed.  We could confirm the sign by working out the mode expansions carefully, but we have seen that~(\ref{negen}) is required by the symmetry of the problem.

In the 1+1 dimensional model, the massless scalar field separates into right-moving and left-moving modes.  More generally, these scatter into one another, and this can be an important effect.  One then has
\be
b_\omega = R_\omega c_\omega + T_\omega \int_{0}^\infty \frac{d\nu}{2\pi}\left( \alpha_{\omega\nu} a_\nu +  \beta_{\omega\nu} a_\nu^\dagger \right)\,. \label{bca}
\ee
Here $c_\omega$ are left-moving modes, coming in from spatial infinity ${\cal I}^-$.  Then $R_\omega$ is the amplitude for them to reflect before reaching the horizon and $T_\omega$ is the transmission amplitude, with $|R_\omega|^2 + |T_\omega|^2 = 1$.  The Hawking flux is then reduced by a greybody factor,
\be
\langle \psi | b_\omega^\dagger b_{\omega'} | \psi \rangle = |T_\omega|^2 \frac{2\pi \delta(\omega-\omega')}{e^{\omega/T_{\rmx H}} - 1} \,. \label{grey}
\ee
Above 1+1 dimensions, the transmission amplitude falls exponentially with the angular momentum~$\ell$.

Some useful orders of magnitude: The typical Hawking quantum has energy  $T_{\rm H} \sim 1/\rs \sim 1/GM$.  The total number of Hawking quanta, and therefore the Bekenstein entropy, is then of order $M/(1/GM) \sim GM^2$; this is the square of the number of Hawking quanta, in Planck units.  The Hawking emission rate is roughly one quantum per light-crossing time $\rs \sim GM$, so the total lifetime is of order $G^2 M^3$.    Factors of 2 and $\pi$ are omitted, and will not be important for our discussion, but can be substantial.  For example, from $T_{\rmx H} = 1/4\pi r_{\rmx s}$ it follows that the wavelength of typical Hawking quanta is $8 \pi^2 \rs$, much larger than the black hole size.

\sect{The information problem}

\subsection{Information loss: pure to mixed}

The first sign of trouble comes if we combine Bekenstein's thought experiment with Hawking radiation.  Imagine that we keep throwing quantum bits into the black hole at a rate such that their energy just equals that of the outgoing radiation.   The black hole's mass and its horizon area stay constant.  However, the nice slice keeps getting longer, and the number of bits on it grows.  So the number of possible states of the black hole grows without bound, and we lose the connection between this and the area.  To recover it, we would need somehow that the bits deep inside can escape with the Hawking radiation, or at least imprint their state on it.   But this is forbidden by causality: once a bit passes through the horizon, it can no longer affect anything on the outside.

This is not yet a crisis.  Maybe we just have to give up the statistical interpretation of the black hole entropy.  But a small refinement of the argument shows that things are more serious.  Suppose we start with a pure state outside the black hole consisting of a large number $n$ of EPR pairs, and as above we throw one of each pair into the black hole.   We end up with a large entanglement,
\be
S_{\rm inside} = S_{\rm outside} = S_{\rm entanglement} = n \ln 2 \,. \label{ent}
\ee
Now we let the evaporation proceed to completion.  The entanglement~(\ref{ent}) cannot decrease, due to causality.  But at the end of the evaporation the black hole disappears, and we are left with half of each pair, in a highly mixed state, with total entropy
\be
S_{\rm entanglement,\, after\,evaporation} = n \ln 2 \,.
\ee
We could have started with the system in a pure state, $S=0$, but it ends up in a highly mixed state. 

This is inconsistent with Schrodinger-like evolution
\be
i \partial_t | \psi \rangle = H | \psi \rangle \,, \label{schrod}
\ee
which otherwise would seem to hold throughout quantum mechanics, quantum field theory, and even string theory.  Similarly the integrated form 
\be
| \psi^{\rm final} \rangle = S | \psi^{\rm initial} \rangle
\ee
or, in a basis,
\be
\psi^{\rm final}_m = S_{mn} \psi^{\rm initial}_n
\ee
takes pure states to pure states.  Instead, black hole evaporation takes an initial pure density matrix
\be
\rho^{\rm initial} = | \psi^{\rm initial} \rangle \langle \psi^{\rm initial} |
\ee
to a mixed density matrix
\be
\rho^{\rm final} = \sum_i p_i  | \psi_i^{\rm final} \rangle \langle \psi_i^{\rm final} | \,.
\ee
In a basis, one has
\be
\rho^{\rmx final}_{mm'} = S \hspace{-7pt} /_{mm',nn'}\rho^{\rmx initial}_{nn'} \,,
\ee 
where $S \hspace{-7pt} /$ is the dollar matrix of~\cite{Hawking:1976ra}.
When the dollar matrix takes the special form 
\be
S \hspace{-7pt} /_{mm',nn'} \stackrel{\rmx Schrod}{=} S_{mn} S_{n'm'}^*  \label{purity}
\ee
with unitary $S$,  this becomes ordinary Schrodinger evolution and takes pure states to pure states, but for a general $S \hspace{-7pt}/$ matrix pure states go to mixed states.


In order to discuss time-dependence, rather than using the frequency modes $b_\omega$ we will often use wavepackets $b_i$.  It is convenient to take the frequency width of the packets to be somewhat narrower than the typical scale $T_{\rmx H}$, so that the width in time is somewhat longer than the light-crossing time $\rs$.  In this case the pairing of $b_i$ and $\tilde b_i$ is approximately diagonal as in the wavefunction~(\ref{bogo}).  This is not essential, but it is convenient for discussion.

We will want to separate the state of the Hawking radiation into quanta emitted before some time, and quanta emitted later, and the quantity that we want to focus on is the van Neumann entropy of the early radiation,
\be
S_{\rmx E} = - {\rm Tr}\, \rho_{\rmx E} \ln \rho_{\rmx E} \,.
\ee
Here $\rho_{\rmx E}$ is the density matrix for the early radiation, tracing out the state of the later radiation and any other degrees of freedom that may be around.  For example, if the whole system is in a pure state $\psi_{mn}$, where $m$ is the state of $\rmx E$ and $n$ is the state of everything else, then
\be
\rho_{{\rmx E}mm'} = \sum_n \psi^*_{mn} \psi_{m'n} \,.
\ee
We are assuming that the modes $b_i$ satisfy a canonical algebra so that the Hilbert space factorizes in this way.  This may break down at some order, but we will see that the effect that we are looking at is robust against small corrections.

The state~(\ref{bogo}) corresponds to a thermal density matrix for a given Hawking photon,
\be
\rho_{mm'}  = \delta_{mm'} p_m\,,\quad p_m = (1 - e^{-\omega/T_{\rmx H}}) e^{-m \omega/T_{\rmx H}} \,,
\ee
in an occupation number basis.  This is a mixed state, with positive von Neumann entropy 
\be
S_{\rm \omega} = - \sum_{m=0}^\infty p_m \ln p_m \,.
\ee
With each Hawking emission, the total von Neumann entropy $S_{\rmx E}$.

\subsection{The Page curve}

There is a nice quantitative and conceptual analysis due to Page~\cite{Page:1993wv}.  This is based on comparing three curves.  The first follows the von Neumann entropy of the Hawking radiation.  As we have discussed, causality requires this to be monotonically increasing, from zero when the black hole forms
 to a value of order the Hawking flux when the system has fully evaporated back to Hawking quanta.  The second curve is the thermodynamic entropy corresponding to the black hole.  It decreases from the Bekenstein entropy of the  original black hole to zero at the end.  Thus we have an increasing and a decreasing curve.  This third curve, that of Page, to good approximation follows the smaller of each of the other two.  That is, it starts at zero, increases to a large value about the middle of the black hole evaporation, and then falls to zero at the end.  All three curves are of order the Bekenstein entropy, of order $G M^2$.  Crudely,
  \be
 {\rm Hawking}:\   0 \to G M^2\,,\quad\quad  {\rm Bekenstein}:\   G M^2 \to 0 \,,\quad\quad  {\rm Page}:\  0 \to G M^2/2  \to 0, \label{curve}
\ee
with the relation to be explained.

Let us assume that the black hole begins in an essentially pure state, for example by collapsing a coherent field.\footnote{As we will show later, the Bekenstein-Hawking entropy is so large that in any normal collapse we can regard the initial entropy of the black hole as negligible.}  
Then the von Neumann entropy of the Hawking radiation at a given time must be equal to that of the remaining black hole state at the same time, and also to the entanglement entropy between the Hawking radiation and the black hole,
\be
S_{\rmx E} = S_{\rmx bh} = S^{\rmx ent}_{\rmx E/bh} \,.
\ee
All three thus follow the same increasing Hawking curve from eq.~(\ref{curve}).

The decreasing curve is the Bekenstein-Hawking entropy.  It should be noted that it is a thermodynamic entropy.  It is different from the microscopic von Neumann entropy $-\rm Tr( \rho\ln \rho)$ (which is probably all that you have encountered at this school so far).  It can be thought of as arising from a coarse-graining of the exact density matrix, and so it can be larger than the true microscopic entropy, but not smaller.  

We see that the curves cross around the midpoint of the life of the black hole, and there we have a problem.  Thus we encounter again the conflict with statistical mechanics noted above.
But things are worse than this.  Let us follow the curves to the end of the life of the black hole.  The Hawking radiation then has a large von Neumann entropy, reflecting its entanglement with the small remaining black hole.  But once the black hole disappears, the Hawking radiation is all there is, and its mixed state is the complete description of the system. 

If the final state of the Hawking radiation is to be pure, then $S_{\rmx E}$ must drop to zero when the black hole disappears.  There is a limit to the rate at which it can decline, and so it must begin to do so about the midpoint of the life of the black hole. This was noted by Page, who also showed that if the black hole dynamics is strongly chaotic, so that the total black hole/Hawking radiation system is in a Haar-random state, then $S_{\rmx E}$ will follow the ascending curve until it nearly meets the descending curve, and then rapidly bend  to follow the descending curve.  This is the Page curve\label{page}.
The point is that the size of a Hilbert space grows exponentially with the number of bits, so that one Hilbert space or the other is much larger, except very close to the crossover.  Whichever system is smaller is then close to a maximally mixed state.  
If the system is not so chaotic, then the actual curve will lie below the Page curve, but in any case it must deviate from the rising Hawking curve no later than the crossover.

A burning piece of coal does follow something like the Page curve.  The early photons are entangled with the remaining coal, but in the end (assuming again that the coal starts in a pure state) the outgoing radiation must be pure.  The burning {\it scrambles} any initial information, making it hard to decode, but it is reversible in principle.  For the coal, the {\it coarse-grained} entropy of the system does follow the Hawking curve.  A common initial reaction to Hawking's claim is that a black hole should be like any other thermal system, and that he had coarse-grained in some way.   But there is a difference: the coal has no horizon.  The early photons from the coal are entangled with excitations inside, but the latter can imprint their quantum state onto later outgoing photons.  With the black hole, the internal excitations are behind the horizon, and cannot influence the state of later photons. 

The difference between the Page curve and the Hawking curve involves subtle physics, in that one must make  complicated measurements of many Hawking quanta to determine which curve one is on, but the difference itself is an order one effect.  Each additional Hawking pair increases the entanglement by an $O(1)$ amount, while we would need to decrease it by an $O(1)$ amount in order to get the information out.  

Mathur nicely sharpened this as follows~\cite{Mathur:2009hf}.  Consider three systems: a Hawking mode $b$, its interior partner $\tilde b$, and all the prior Hawking radiation $E$.  For a Hilbert space with three factors, the von Neumann entropies satisfy strong subadditivity, here
\be
S_{\tilde b b} + S_{bE} \geq S_b + S_{\tilde b b E}\,. \label{mathur}
\ee
Now, $\tilde b b$ is in a pure state, the $a$-vacuum, so $S_{\tilde b b} = 0$ and $S_{\tilde b b E} = S_E$, so this becomes
\be
S_{bE} - S_{E}\geq S_b   \,.  \label{ssa2}
\ee
(In fact, equality must hold, by ordinary subadditivity.)
But this is saying that we are on the Hawking curve: the von Neumann entropy of the radiation after emission of $b$, minus that before emission, is just equal to the von Neumann entropy of $b$ by itself.  In order to follow the Page curve, we would need the right side to be something of the same order but negative.\footnote{Including the greybody effect~(\ref{bca}) reduces the magnitude of the discrepancy but does not eliminate it.}  Eq.~(\ref{ssa2}) says that there is no entanglement between $E$ and $b$, while the Page curve requires that they be entangled.  An important aspect of this argument is that the only degree of freedom behind the horizon that appears is $\tilde b$.  Many discussions of information loss use the state on the entire long nice slice in the interior, as we have done above, but Mathur's argument shows that this is not needed, the problem is already present right in the neighborhood of the horizon.

We can understand the point without the full machinery of strong subadditivity (which is apparently challenging to prove).  The Hawking process produces $b\tilde b$ in a pure state.  The wavefunction must then be a product $\psi_{\tilde b b} \psi_{\rm E}$.  But this allows no entanglement (or, more precisely, mutual information) between $b$ and $E$.\footnote{This omits the greybody factors.  I think that one can include their effect just by extending $\tilde b b$ to $\tilde b b c$.}

Another possibility is that the black hole does not completely evaporate.  Once the black hole mass and radius approach the Planck scale, we no longer trust the low energy calculation of Hawking radiation, and perhaps the evaporation stops.  In this case the system would end in a pure state, where a Planck scale remnant with a large number of internal states would be highly entangled with the Hawking radiation.  The remnant might be stable or simply very long-lived.  In the latter case, the quantum state can be encoded in very low-energy photons emitted over a very long time scale (since the amount of energy available in the remnant is small).

\subsection{The three classical alternatives}

Thus, there are three broad possibilities: (1) that information is lost in the manner that Hawking argued, (2) that the information is carried by the Hawking radiation, as it would be for an ordinary thermal object, or (3) a remnant of some sort.  Most scenarios end up in one of these three classes, though sometimes in a non-obvious way. 
Each of these possibilities has positive and negative features.  

{\it Information loss}, the $S \hspace{-7pt}/$ matrix, seems to follow logically from treating gravity as an effective field theory.  But if pure states can evolve to mixed in black hole decay, then the basic framework of quantum mechanics is changed, and one would expect this effect to appear elsewhere.  Quantum gravity effects may be significant only at the Planck scale, but general principles of quantum field theory imply that this will feed down to low energy through low dimension operators, and there are strong limits on such effects~\cite{Ellis:1983jz}.  Even worse, it has been argued that $S \hspace{-7pt}/$-matrix evolution leads to strong violation of energy conservation~\cite{Banks:1983by}.  Time translation invariance implies that $S \hspace{-7pt} /_{mm',nn'}$ is nonvanishing only for $E_m - E_n - E_{m'} + E_{n'} = 0$.  When the evolution is unitary~(\ref{purity}),
both $E_m - E_{n}$ and $E_{m'} - E_{n'}$ vanish, but the $S \hspace{-7pt}/$-matrix generically contains only the single $\delta$-function.  Moreover, if the purity-violating effects are local in time they should be polynomial in energy, making a second $\delta$-function impossible (this is a paraphrase of~\cite{Banks:1983by}).
It follows that a density matrix of definite energy, $E_n =  E_{n'}$, can evolve to one of different energy,  $E_m =  E_{m'} \neq E_n$.
  If one follows the usual effective field theory rule of parameterizing all Planck scale effects by operators of appropriate dimension, but now allowing information loss, then these will violate energy conservation in a large way.

{\it Information conservation} via the Hawking radiation is attractive for the universality of thermal behavior.  However, it seems to require superluminal transport of information from the black hole interior.  We have been studying quantum gravity as an effective field theory, and this has led us to information loss or possibly remnants.  To get the information out  with the Hawking radiation, it seems that effective field theory must break down even in regions where curvatures are small.  Note that the details of the UV theory have not entered.  We have made one implicit assumption about the UV, namely that the adiabatic theorem applies, so that when the Hawking modes-to-be redshift down into the effective field theory they are empty.  But as Andrei Linde points out, this redshifting is happening all around us due to the expansion of the universe, and if the adiabatic principle did not hold we would immediately fry. 

{\it Remnants}, whether stable or long-lived, give up the connection between Bekenstein-Hawking entropy and density of states (so also does information loss).  They also seem  problematic for other reasons.  We could have started with an arbitrarily large black hole, so the number of states that must be available to this Planck-sized remnant is unbounded above.  If follows that thermal equilibrium cannot actually exist, and even at zero temperature the virtual effects due to remnants would seem to diverge, as would remnant pair production amplitudes.
 
Most ideas fit into one of these categories, though not always in an obvious way.  The discussions become rather abstruse, but let me mention a couple of examples that may come to mind.   Late in the black hole lifetime, near the singularity, low energy effective field theory breaks down.  One scenario that has been considered is that the true geometry fills in the region after the singularity in such a way that causality allows the internal degrees of freedom to escape.  This would be a `long-lived remnant.'  The outgoing radiation cannot change until the black hole radius is nearly Planckian, so we are well past the crossing of the entropy curves.   

Instead, one might imagine that the singularity is replaced by a baby universe.  The baby universe and the external Hawking radiation-fill universe are disconnected but highly entangled.  From the point of view of the disconnected exterior, the quantum theory describes a system of large entropy, so information is lost, with the consequences such as energy nonconservation as discussed.  The interpretation from the global point of  view, with the baby universe, is  more subtle~\cite{Strominger:1994ey}.  If one sums coherently over all the ways this can happen, then information will not be lost but rather the black hole $S$-matrix will depend on the baby universe wavefunction (a.k.a. the $\alpha$-parameters).  The first black holes to evaporate will have an unpredictable final state, but once the wavefunction of the baby universe is measured, the remaining decays become predictable.  In the end this again becomes a long-lived remnant.  It also seems that attempts to avoid the energy nonconservation with information loss lead to a similar picture~\cite{Oppenheim:2009nb,Unruh:2012vd}.

\sect{AdS/CFT}

\subsection{Ideology}

One could go around in circles indefinitely, debating the pros and cons of the various alternatives above.  Fortunately, insight came from a new direction, AdS/CFT duality.  The gist of the argument is that the formation and decay of a black hole is dual to some process in the gauge theory, and in the gauge theory pure states evolve to pure states.  Actually, this argument was already made for the BFKS Matrix theory, which is  part of the broader set of gauge/gravity duals, but the issues are somewhat clearer in AdS/CFT.
So most of this section will be devoted to fleshing out this argument, but along the way I will develop some useful general facts about black holes in AdS, and about black holes and statistical mechanics in general.

Joao gave a nice introduction to {\rm AdS/CFT}.  As with almost all discussions of gauge/gravity duality, it is assumed that the nonperturbative CFT is well-defined, and all properties of the gravitational theory are derived from this.  Having a nonperturbative construct of the CFT is very powerful.  This gives a good description of the gravitational theory in the region of long-distance gravity.  The Page curve, which holds in any normal quantum theory such as the CFT, then excludes remnants and information loss and so implies purity of Hawking radiation.  

An important question to keep in mind is whether there is an independent nonperturbative construction of the gravitational theory, as there is with other dualities.  Another, which must eventually be faced, is how to describe quantum gravity in more general backgrounds, especially cosmological.  The black hole problem gives us a more immediate challenge: AdS/CFT tells us what happens to the information, but not how.
Understanding this my give us clues to those deep question, as with the original duality.


The canonical example of a nonperturbative CFT with a low energy gravitational dual is
\be
{\cal N} = 4\,, d=4\,, \mbox{$SU(N)$ Yang-Mills}\ \Leftrightarrow\ \mbox{IIB string theory in asymptotic $AdS_5 \times S^5$} \,. \label{adscft}
\ee
This is `derived' (I still like my 2010 TASI lectures here~\cite{Polchinski:2010hw}) by assuming that the weak-strong interpolation between D-branes and black branes commutes with the low energy limit on the two sides.  This conclusion is supported by a great deal of evidence of various sorts, but it is important to ask, is the duality sufficiently exact that we can make this argument, especially since the difference between a pure state and a mixed state involves rather subtle measurements?  We believe that the duality is sufficiently constrained to imply this.

The duality involves two parameters $N$ and $\lambda = g_{\rm YM}^2N$.  In the string description there are three key length scales, the Planck length $l_{\rm P}$, the string length $l_{\rm s}$, and the curvature radius $l_{\rm AdS}$.  These are related to the rank $N$ of the gauge group and the gauge coupling by
\be
\frac{l_{\rm AdS}^4}{l_{\rm P}^4} \sim N \,,  \quad \frac{l_{\rm AdS}^4}{l_{\rm s}^4} \sim \lambda  \,. \label{ratio}
\ee
A classical gravity description emerges only when both are large.  One might imagine a weak version of the duality, which might only be asymptotic in one or both parameters, so that information loss might exist but not be visible in the asymptotic expansion.  

In fact, the general assumption is that the duality is exact at all values of $N$ and $\lambda$, that there is essentially a unique quantum theory for given CFT parameters.  Quantum theories are highly constrained, and it is hard to see how the theories on the two sides of the duality could have so many points of agreement without being exactly the same.  In the case of explicit examples like~(\ref{adscft}) we can identify in the gauge theory a complete set of supergravity states in $AdS_5 \times S^5$, and their trilinear couplings agree with supergravity, including the coupling of the graviton to energy --- this is guaranteed by supersymmetry.  So we can say that this is {\it some} theory of quantum gravity, and  I will take the point of view that the CFT provides the precise definition the theory.  And there is abundant additional evidence, such as the existence of various stringy states, to believe that it is IIB string theory.  Most important for us, the gauge theory contains states with the right properties to be AdS black holes.

We will be using global AdS$\ \times\  S^5$.  The  metric is
\be
ds^2 = -(1 + r^2/l_{\rm AdS}^2) dt^2 + \frac{ dr^2}{1 + r^2/l_{\rm AdS}^2} + r^2 d\Omega^2_{S^3} + l_{\rm AdS}^2 d\Omega^2_{S^5} \,. \label{adsmet}
\ee
The duality implies a one-to-one matching of the Hilbert spaces.  The IIB string is quantized such that the metric approaches~(\ref{adsmet}) at the boundary $r \to \infty$.  The gauge theory is quantized on $S^3\ \times$ time.  There is also a matching of the observables.  The simplest are local operators in the CFT.  Each CFT operator ${\cal O}_i$ has a corresponding bulk field $\phi_i$.  The insertion of ${\cal O}_i$  in the CFT path integral is equivalent to a perturbation of the boundary condition on $\phi^i$, or equivalently an insertion of the operator $\phi_i$ at the boundary,
\be
\lim_{r \to \infty} r^{\Delta_i} \phi_i(r, t, \theta) = {\cal O}_i(t,\theta) \,.  \label{extrap}
\ee
The angles are those of the gauge theory $S^3$.  The angular dependence on the $S^5$ is encoded in the dependence of ${\cal O}_i$ on the scalar fields, which is encoded in the index $i$.

Let us be a bit more precise about the Hilbert space.  In QFT, one expects to be able to reach any state by acting on the vacuum with a product of local operators at various times; in fact, this is an axiom in constructive quantum field theory.  (In some cases there could also be topological sectors, but this will not be central.)  The dictionary~(\ref{extrap}) implies that each operator creates or destroys a bulk excitation, and the bulk Hilbert space consists of states that could be created from the vacuum in this way.  By throwing in enough energy, we could create a black hole, and  we will be studying black hole states that could be created in collapse in this way.  

One might wonder whether an understanding of AdS black holes is enough.  If $l_{\rm AdS}$ is large enough, the curvature of the background will be negligible in comparison to the size of the black hole.  Hawking's calculation and paradox then go through just as in flat spacetime, so AdS is a good arena for studying this.  Of course, putting a physical system in a box is often useful in making things well-defined. 

 Indeed, in infinite flat space one might wonder if there really is a paradox.  Perhaps it is not possible to capture and manipulate all the outgoing quanta with sufficient precision to distinguish a pure state from a mixed state.  Maldacena noted a paradox in AdS which is presumably equivalent, but simpler to frame~\cite{Maldacena:2001kr}.  Consider a two-point function in some black hole state,
\be
\langle \psi | {\cal O}_i(t,\theta)  {\cal O}_i(0,\theta') | \psi \rangle \,. \label{tpf}
\ee
In a bulk calculation, the excitation created at time zero will rattle around, bouncing off the boundary, with some amplitude to fall through the horizon.  In low energy effective theory, once it falls behind the horizon it is trapped.  This process continues indefinitely, so the two-point function goes to zero over long times.  However, the gauge theory on the compact space $S^3$ has a discrete spectrum.  By inserting a completes set of 
states,
\be
\sum_{a,b} e^{it(E_a - E_b)} \langle \psi |a \rangle \langle a| {\cal O}_i(0,\theta)  |b \rangle \langle b| {\cal O}_i(0,\theta)|  \psi \rangle \,, 
\ee
one obtains order $e^{2S}$ states each of magnitude $e^{-2S}$, where $S$ is the black hole entropy.  Over long terms this averages to of order $e^{-S}$, rather than falling to zero.  This may not sound like the same paradox, but again it is a conflict between the low energy field theory and the exact quantum theory.  It only requires measuring a two-point function, but one must make many measurements because one is looking for a small answer $O(e^{-S})$.  To put this another way, we would like to be able to see from the gravitational bulk theory that the black hole spectrum is discrete: this would be equivalent to solving the Hawking paradox.

\subsection{Black holes in AdS}

There are actually three kinds of black holes in AdS spacetime~\cite{Horowitz:1999uv}.  To begin, it is interesting to compare 
the entropy of a black hole with that of ordinary matter of the same mass and volume.  I will do this first for black holes in $D=4$.  
For a relativistic gas in a volume of typical length $l$,
\be
M \sim T^4 l^3 \,,\quad S_{\rmx gas} \sim T^3 l^3\ \Rightarrow\ S_{\rmx gas} \sim (Ml)^{3/4} \,.
\ee
I am omitting here numerical factors
.
For  $l$ of order the black hole radius $\rs$,
\be
S_{\rmx gas} \sim (M\rs)^{3/4} \sim (M\lp)^{3/2} \sim (A/\lp^2)^{3/4} \,.
\ee
For a larger-than-Planckian $\rs$ the black hole entropy is always larger than that of the gas.  If we try to form a black hole by collapsing a gas of ordinary matter, we can only populate some small subset of the black hole Hilbert space,
\be
e^{O(A/\lp^2)^{3/4}} \lll e^{O(A/\lp^2)}\,.
\ee
To form a general state one would have to throw in matter over a longer time $t_{\rm bh}$, essentially running the black hole evaporation in advance.  This is also a motivation for the holographic principle: in a given region, the greatest number of states is obtained by filling it with a large black hole, whose entropy goes as the surface area of the region.  

In general $D$ we similarly get 
\be
S_{\rmx gas} \sim (Ml)^{(D-1)/D} \,. \label{gass}
\ee
for a gas of energy $M$ in a volume of length $l$.  Let us also extend our earlier calculation of the black hole lifetime to general $D$.
The units of the gravitational coupling $G_D$ are $l^{D-2}$, so the Planck length is defined $G \sim l_{\rmx P}^{D-2}$.
The Newtonian potential is of order $G_D M / r^{D-3}$, and setting this to 1 gives the magnitude of $\rs$,
\be
 \rs^{D-3} \sim G_D M\,.
\ee
The Hawking temperature is $1/\rs$, while the BH entropy is the corresponding area,
\be
T_{\rmx H}\sim 1/\rs\,,\quad   S_{\rmx BH} \sim  \rs^{D-2}/G_D \,, \label{bhs}
\ee
and so the total lifetime is 
\be
t_{\rm bh} \sim \rs S_{\rmx BH} \sim  \rs^{D-1}/G_D \,.
\ee

Let us now ask what happens as we vary the mass (which we will write in terms of $\rs$) with fixed $G_D$ and $l_{\rm AdS}$.  First, for 
\be
\rs <  l_{\rm AdS} \left(\frac{l_{\rmx P}}{l_{\rm AdS} }\right)^{(D-2)/(D-1)} \,,
\ee
then $t_{\rm bh} < l_{\rm AdS}$ and the black hole evaporates so quickly that the evaporation is completed before the decay products 
even travel a distance of order the AdS radius. So the AdS box does not have much effect, and the black hole is unstable as in flat spacetime.  Note that the relevant $D$ here is 10, because the $S^5$ and $AdS_5$ are both large.

In fact, as we increase $\rs$, as long as 
\be
\rs <  l_{\rm AdS} \left(\frac{l_{\rmx P}}{l_{\rm AdS} }\right)^{(D-2)/(2D-3)} \,,
\ee
the gas entropy~(\ref{gass}) exceeds the black hole entropy~(\ref{bhs}) and the black hole is again unstable to decay.  Using ${l_{\rmx P}}/l_{\rm AdS} \sim N^{-1/4}$ and $D = 10$ this becomes
\be
\rs <  l_{\rm AdS} N^{-2/17} \,.
\ee
For
\be
l_{\rm AdS} N^{-2/17} < \rs <  l_{\rm AdS} \,,
\ee
the black hole is stable.  It is still much smaller than the AdS length so is effectively living in 10-dimensional flat spacetime.  Finally, for 
\be
\rs > l_{\rmx AdS}\,,
\ee
the black hole no longer sits in an approximately flat 10-dimensional space.  It undergoes a transition to a so-called big black hole, which is uniform on the $S^5$:
\be
ds^2 = -(1 + r^2/l_{\rm AdS}^2 - \mu/r^2) dt^2 + \frac{ dr^2}{1 + r^2/l_{\rm AdS}^2 -\mu/r^2} + r^2 d\Omega^2_{S^3} + l_{\rm AdS}^2 d\Omega^2_{S^5} \,. \label{bbh}
\ee
At larger masses,
\be
\rs \sim M l_{\rm AdS}^{4-D}  \lp^{D-2} \,,\quad T_{\rmx H} \sim \rs/\la^2 \,.
\ee
The temperature-radius relation is standard holography.

For $AdS_5 \times S^5$ we thus have three phases at given increasing energies: a gas, a 10-dimensional black hole, and a 5-dimensional black hole.  If you work out the temperature $T = dM/dS$, you find that it increases with $M$ in the gas phase and the 10-d phase, and falls in the 5-d phase (negative specific heat).  The intermediate phase is stable in the microcanonical ensemble (fixed $M$) but not the canonical ensemble (fixed $T$): if we put the system in contact with a heat bath, it will decay to one of the other two phases, and the equilibrium state is given by a Maxwell construction.  

Thus, as we increase $T$, there is a single transition, from the gas to the 5-d black hole.\footnote{How can their be a transition temperature in a conformal theory?  The point is that global AdS is dual to a CFT on a sphere, and the radius of the sphere sets the scale.}
In the gas phase, the entropy is 
\be
S \sim (\la T )^{D-1} \,.
\ee
In the 5-d black hole phase it is
\be
S \sim \rs^3 \la^5 /\lp^8 \sim (\la T )^3 N^2\,,
\ee
where the middle expression is the product of the $S^5$ area with the area of an $S^3$ of radius $\rs$.  The $N$-dependence is important: this corresponds to a deconfinement transition.\footnote{It is dimensionfully odd to have a  bulk length $\la$ appear in a CFT expression.  This happens because $-g_{\Omega\Omega} / g_{tt} \to  \la^2$ in the metric~(\ref{adsmet}) at large $r$, so the radius of the CFT $S^3$ is equal to the bulk scale $\la$.}  One expects this same transition in the gauge theory, where it takes the form of a deconfinement transition: at low energy one has color singlets, and at high energy adjoints, hence the powers of $N$.  This transition is even visible at small $\lambda$, where one can calculate in the CFT.  See~\cite{Witten:1998zw} for more discussion and both bulk and CFT references.

Finally we can run our thought experiment.  The most direct way to do this is to inject a few high energy particles with total energy less than the transition to a stable black hole.  They will have some amplitude to collide and form a dynamical black hole, which will then decay into ordinary quanta.  This CFT dynamics takes pure states to pure states, so we can conclude that information is not lost.  Sometimes it is more convenient to work with the stable black holes, since these represent true equilibrium states of the CFT.  To run the same thought experiment there, we couple the CFT to another field theory with many more degrees of freedom, so that energy can escape.  Also, by controlling the coupling, we can consider a variety of new thought experiments.

We have argued against information loss, but what about remnants?  Recall that for this to solve the problem, we would need an unbounded number of low energy states.  The CFT is strongly coupled so we cannot be sure about its dynamics, but this goes against all our experience with gauge theories.\footnote{The most obvious place to look for a lot of states is far out in scalar field space, but there the gauge symmetry is broken to $U(1)$'s and we can reliably say that the density of states is bounded.}  Note that for the purposes  of studying the paradox, we need AdS to be large but not infinite in size.  One might imagine some infinite density of states coming in $N$ and $\lambda$ go to infinity, but we do not need to consider with this.  So it seems that we exclude remnants as well.  Note that this depends on our knowing something about the nature of the dual CFT. 

\subsection{More on black holes and statistical  mechanics}

I would like to dispell a possible confusion, about the relation between pure and mixed states in statistical mechanics.  This is not directly related to the information paradox, but is good to be clear about.  

Let us start by considering two sealed containers, each with a chunk of ice at 0K.  We put the first container A in contact with a heat bath at 400K until equilibrium is reached and the container is full of steam, and then decouple the heat bath.  We heat the second container B with a laser until the same energy density is reached, and after it comes to equilibrium we again have a container full of steam.  Is there any difference between the two?  The first is in the usual mixed state (ensemble): it is entangled with the bath.  By itself it is described by a density matrix
\be
\rho_A = \frac{e^{-\beta H_A}}{{\rm Tr}(e^{-\beta H_A}) }\,.
\ee
The second is in some pure state,
\be
\rho_B = | \psi \rangle\langle \psi | \,.
\ee
I give you both containers.  Can you tell which is which?

If I tell you exactly how I prepared B, then indeed you can tell by applying the projection operator $ | \psi \rangle\langle \psi |$, giving the identity for B but essentially zero for A.  But if I only tell you that I have prepared a pure state, but not how, then you cannot tell.  This is because, first, as you make measurements on A you `collapse' it into a random pure state, and, second, it is a property of complex (chaotic) systems that all pure states look essentially the same (the eigenstate thermalization hypothesis is one precise statement in this direction~\cite{ETH}).  Note that these two systems have the same thermodynamic entropy, but very different von Neumann entropies: the first is equal to the thermodynamic entropy, and the second is zero.

Now let us heat both even more, until the steam collapses into a black hole (or just use AdS/CFT duality to get to a dual black hole picture).  Now the same question, and the same answer: we cannot tell the difference.  So black holes can be in pure states.  Indeed, we implicitly assumed this when we talked about the Page curve.

However, if I prepare $2n$ copies of A, or $2n$ copies of B, then you can tell which with high fidelity~\cite{Hayden:2007cs,Harlow:2014yka}).  Consider the swap operator $Z$, which just switches two containers:
\be
Z |\psi_1, \psi_2 \rangle =  |\psi_2, \psi_1\rangle \,.
\ee
Now, $Z$ has eigenvalues $\pm 1$, and 
\be
\langle \psi_1, \psi_2 | Z |\psi_1, \psi_2 \rangle = |\langle \psi_1 |\psi_2 \rangle|^2  \,.
\ee
So in case B we get +1 with certainty.  But in case A, for all but a negligible part of the ensemble the two containers are in different states and we get a mean of 0, meaning $\pm 1$ with equal probability.  So if we measure $Z$ and get $-1$, we know that we have A, while if we get $+1$ we can't be sure.  But if we do this with $n$ pairs and get $+1$ each time, we know with probability $1 - 2^{-n}$ that we have B.\footnote{Another thing we could do with multiple copies is to measure tiny $O(e^{-S/2})$ fluctuations from the mean ensemble behavior, which will be repeatable in case B.  But in order to get statistical significance we would need $e^S$ copies.}

By the same measurement we can tell whether black holes destroy information.  Prepare $2n$ initial copies of the identical pre-collapse state, and then collect the Hawking radiation and do the swap test.  For coal we get B.  For the black hole, Hawking in 1976 would say A, and AdS/CFT would say B.

You set up an interference experiment, and if any pair interfere then they are in the same pure state.  This is the simplest way to distinguish pure from mixed states.  If you wanted to design an experiment to see whether black holes destroy information, this is the simplest way (see the discussion of the `swap test' in \cite{Hayden:2007cs,Harlow:2014yka}).  By the way, this kind of measurement is being done in the lab, at the level of systems of four cold atoms~\cite{rislam}.

This is standard QM,  for observations of the black hole from the outside.  For the inside, who knows?  
That is the subject of the remaining lectures.

\sect{Black hole complementarity and its failure}

If we believe AdS/CFT, then the information escapes with the Hawking radiation.  How, then, is Hawking's argument evaded?  How does information travel from inside the black hole to the outside?  AdS/CFT is a bit of a black box here.


Black hole complementarity (BHC) says, in a sense, that there was never a paradox at all~\cite{Susskind:1993if,Stephens:1993an,Susskind:1993mu,Preskill}.  Let us start outside the black hole with an entangled pair of bits
$q_1,q_2$, and throw $q_2$ in.  Information conservation implies that the state of $q_2$ is eventually carried by the later Hawking radiation: this bit $q_3$ encoded in the later radiation is also entangled with $q_1$.  The problem is that there is a nice slice containing all three bits $q_1, q_2, q_3$, so strong subadditivity is violated for the state on this slice.  That is, $q_1$ and $q_2$ are together in a pure state, so $q_1$ cannot be entangled with the later emitted mode $q_3$.

However, it is difficult for any single observer to see both copies $q_2, q_3$ of the seemingly identical state.  They would have to wait for the copy $q_3$ to emerge, measure it, and then jump in and see the original $q_2$ before it hits the singularity (not the reverse, which would be causal).  Even with an assistant jumping in earlier and redirecting $q_2$ to avoid the singularity as long as possible, the exponential redshifting allows only a time of order $\rs \ln (\rs/\lp)$ to see $q_2$ (a more careful treatment suggests $2 \rs \ln (\rs/\lp)$).  This is known as the scrambling time~\cite{Hayden:2007cs,Sekino:2008he}.  It is very short compared with the black hole lifetime $\sim \rs  (\rs/\lp)^{D-2}$, so only a small delay in the emission time for the information avoids any conflict.


Focusing on what can be measured operationally was important in formulating special relativity, general relativity, and quantum mechanics, and so we should take such limitations seriously.  Still, it is some distance from the uncertainty principle to the wavefunction and its interpretation.  If the Hawking radiation carries information away, low energy effective field theory must break down prematurely, but in what way?  BHC gives a hint.  It says that the framework of the theory must be modified in a fundamental way: the wavefunction on a spacelike slice like the nice slice, which is the basic object in canonically quantized gravity, is not a part of the exact theory.  Rather, we should restrict to parts of the slice that a single observer can see.

One possible interpretation of this is that there is still a global Hilbert space, but that operators in different spatial patches which would ordinarily commute do not do so.  So observers in different patches are seeing the same bit in a single Hilbert space, but they are interpreting it in different ways.  In this approach, let us call it global complementarity, the ordinary structure of quantum physics is preserved, but locality breaks down in an extreme way.  This seems morally consistent with AdS/CFT, where there is a global Hilbert space, but in one interpretation the physics lives in the boundary and in another it lives in the bulk.  

Another notion, which has come to be called strong complementarity, is that there is no global Hilbert space, but rather each observer has their own, with some matching condition on the overlaps.  This is already a weakening of quantum theory, because the Hilbert space structure is subordinate to some causal structure.   Further we can ask whether the matching condition, in strong complementarity, or the structure of observables, in global complementarity, respect the usual structure of quantum mechanics.  The arguments that we will discuss in the rest of these lectures call all of these into question, and there may be no consistent theory of the black hole interior: it may not exist.

The essence of black hole complementarity is that information is not lost, but that no single observer sees anything funny.  This was enshrined as a set of postulates for BHC:

\begin{quote}

{Postulate 1 (Purity):} {The process of formation and evaporation of a black hole, as viewed by a distant observer, can be described entirely within the context of standard quantum theory. In particular, there exists a unitary $S$-matrix which describes the evolution from infalling matter to outgoing Hawking-like radiation.}

{Postulate 2 (Effective field theory, EFT):} {Outside the stretched horizon\footnote{We will not try to define this, simply taking it to be near-Planckian}
 of a massive black hole, physics can be described to good approximation by 
gravity as an effective field theory.}

{Postulate 3 (Microscopic BH entropy):} To a distant observer, a black hole appears to be a quantum system with discrete energy levels. The dimension of the subspace of states describing a black hole of mass $M$ is the exponential of the Bekenstein entropy $S(M)$.

{Postulate 4 (No drama):} A freely falling observer experiences nothing out of the ordinary when crossing the horizon.  `Out of the ordinary' would include high energy particles in violation of the adiabatic principle (fiery drama), and also measurements that could not be described within the ordinary framework of QM (quantum drama). 
\end{quote}
We are going to argue that these postulates are inconsistent.  But no-go theorems are abstract and tricky.  It is better to start by trying to make a model of what these postulates mean, and only after failing use it to motivate a no-go result.

A model of these postulates that one might keep in mind is the membrane, or stretched horizon, paradigm.  In this picture, the black hole  as seen from the outside can be regarded as a dynamical membrane just outside the horizon, which absorbs information, scrambles it, and reemits it.  All the nonperturbative dynamics is confined to this membrane, with ordinary effective QFT outside it (postulate 2).  On the other hand, the infalling observer passes through the horizon smoothly, just as Einstein's theory would predict.  Seen from the outside, the bit $b_2$ is absorbed by the horizon, resides there for a while, and is eventually reemitted.  
This is an appealing picture, but it seems that the actual situation must be a good deal stranger.

Another toy model of BHC that may be useful is the `bit model'~\cite{Mathur:2009hf,Giddings:2011ks}.
Model the Hawking radiation by a line of quantum bits, each of which can take the value 0 or 1:
\be
| {\pmb b} \rangle = |\tilde b_{1} \tilde b_{2} \ldots \tilde b_{n} | b_n  b_{n-1} \ldots b_1\rangle \,. \label{2nb}
\ee
The central vertical bar divides the bits inside, on the left, from the bits outside. The black hole wavefunction is some superposition of these $2^{2n}$ states,
\be
\sum_{\pmb b} \psi_{\pmb b} | {\pmb b} \rangle 
\ee
One can think of these bits as the right-movers, the functions of $U$.  We could make things more elaborate by adding left-movers if needed.

Now consider how this evolves in time.  The bits are described in the infalling coordinate $b$.   Due to the gravitational redshift, as time goes on these bits move to larger values of the black hole translation coordinate $a$: this energy increases.  In order to keep the same energy scale, one must then add a pair of bits at each time step,
\be
| {\pmb b} \rangle = |\tilde b_{1} \tilde b_{2} \ldots \tilde b_{n}\tilde b_{n+1} | b_{n+1}   b_n  b_{n-1} \ldots b_1\rangle \,. 
\ee

In the Hawking process, the model evolves
\be
| {\pmb b} \rangle \to c_0 |\tilde b_{1} \tilde b_{2} \ldots \tilde b_{n} \tilde 0 | 0 b_n  b_{n-1} \ldots b_1\rangle +  c_1|\tilde b_{1} \tilde b_{2} \ldots \tilde b_{n} \tilde 1 | 1 b_n  b_{n-1} \ldots b_1\rangle \,. \label{bmhawk}
\ee
This is a schematic version of the Hawking process: roughly once per time $\rs$, the black hole emits a Hawking quantum to the outside, and the nice slice inside becomes one bit longer.  The new pair appears in the entangled state $c_0 |\tilde 00\rangle + c_1 |\tilde 11\rangle)$, analogous to the squeezed state~(\ref{bogo}).  The Hilbert space is larger by a factor of 4, but the old Hilbert space maps into the new one.

The entanglement entropy increases monotonically, with the undesirable consequences discussed earlier.  What we want is that, once the Page curve turns over, the interior Hilbert space becomes one bit smaller.  One can use the bit model to explore alternate evolutions~\cite{Mathur:2009hf,Giddings:2011ks}.
One such model would be
\be
| {\pmb b} \rangle \to | \tilde b_3 \ldots \tilde b_{n} \tilde b_1 | \tilde b_2 b_n  b_{n-1} \ldots b_1\rangle  \,.  \label{bmunit}
\ee
Here, the two left-most  bits are removed from the left and inserted in place of the previous 
$\tilde b_{n+1} | b_{n+1}$.  This has the desired effect on the size of the Hilbert space, but now $\tilde b_1|   
\tilde b_2$ do not appear in the $a$-vacuum so we have a problem.

In the Hawking evolution~(\ref{bmhawk}) the new bit $b_{n+1}$ on the outside is entangled with its partner $\tilde b_{n+1}$ in the interior.  In the unitary evolution~(\ref{bmunit}), the new bit $\tilde b_2$ on the outside is entangled with the earlier Hawking bit $b_2$.  One could make this more realistic by including a unitary operation on the interior Hilbert space at each step, reflecting the expected chaotic dynamics of thermal systems.  In this case, the new bit would be entangled in a more complicated way with the exterior, but this would not change the result.

Thus the bit model does not allow us to evade the Hawking problem, but gives us a way to parameterize it: information loss in the model~(\ref{bmhawk}) and modification of the horizon behavior in the model~(\ref{bmunit}).  The latter model is a bit coarse in scale, and so can be realized in different ways as we will see later, including modification of the geometry at large distances~\cite{Giddings:2012gc} or the firewall
\cite{Almheiri:2013hfa}.

The bit model does not incorporate black hole complementarity: no observer can see all the bits~(\ref{2nb}).  We could try to make some sort of complementary bit model, either in the context of global complementarity or strong complementarity.  This was the immediate origin of the firewall argument, but Mathur's subadditivity argument shows that this is no help: there is a contradiction already at the level of the subsystem $\tilde bbE$, because a single observer can see all of these bits, as we will see.  This includes only one bit behind the horizon, and is sufficient  to describe the quantum mechanics of a single observer who interacts with the Hawking radiation for a period of time and then jumps into the black hole.

In summary, we have an order one problem at the level of the two-point function that measures the bits $\tilde b_{n+1} | b_{n+1}$.  Mathur express this as follows: if information is carried by Hawking radiation, then the black hole horizon cannot be information-free (well-described by low energy effective field theory).  The bit model is rather coarse, its resolution in time and space is of order $\rs$.  In order to see what might be happening, we will return to the full black hole system.

Consider the postulates of BHC, focusing on a late Hawking mode $b$, emitted during the period when the Page entanglement entropy is decreasing.  For reference to Mathur's earlier discussion~(\ref{mathur}), the three bits $b,b',E$ can the though of as a three-dimensional slice through the bit model.

Postulate~1, purity of the Hawking radiation, implies as we have seen that $b$ must be entangled with $E$.  Postulate~2 we use to argue that the mode $b$ propagates from the near-horizon region to the asymptotic region according to low energy effective field theory; since we are dealing with an order-one problem, we only need this to order one accuracy.  Postulate~3, the statistical interpretation of the Bekenstein-Hawking entropy, we do not actually need; postulate~1 already implies that the Page entropy has to drop to zero at the end of the decay, and in order for the Hawking radiation to carry the entanglement, the decline must begin around the midpoint of the decay, if not before.  But it largely goes hand-in-hand with postulate~1, so we will take it for granted.  Postulate~4, no drama, requires that $b$ be in a pure state with $\tilde b$ (or with $\tilde b c$ to account for greybody factors).  Further, it requires that the infalling observer, who has access to all of $E$, $b$, $\tilde b$, has a normal quantum mechanical description.  But subadditivity says that the entanglements required by postulate 1 and postulate 4 are inconsistent with QM for the infalling observer. 

It is interesting to contrast this with the thought experiment used to justify BHC.  There, no single observer could see the cloned bits $q_2$ and $q_3$, each of which is entangled with $q_1$.  Here the cloned bits are $\tilde b$ and $b_E$, and the difference is that $b_E$ comes out much earlier.

Other versions of the paradox have been given~\cite{Almheiri:2013hfa,Bousso:2013wia,Marolf:2013dba}.  These use the same basic assumptions, though details differ.  For example, some apply to typical black hole pure states, as opposed to the highly-entangled late-time evaporating black hole.

\sect{So, what to give up?}

\subsection{Violation of purity}

Purity, which explicitly assumes that the information is carried by the Hawking radiation, would be violated by either information loss or by remnants.  Either of these would mean that the entropy does not follow the Page curve but rather continues to rise according to Hawking.  There is then no requirement for $E$ and $b$ to be entangled, and so no contradiction.

I have discussed the classic arguments against information loss and remnants, and the further evidence against from \mbox{AdS/CFT}.  
I continue to find these persuasive, though those who never gave up on these ideas, e.g.~\cite{Unruh:1995gn,Chen:2014jwq}, are saying, ``we told you so!''  But I see AdS/CFT as forward progress, as opposed to trying to cobble together some consistent version of quantized metrics.

\subsection{EFT/NVNL}

If we give up EFT, we can fix the conflict if, as $b$ propagates from the region near the horizon out to several times $\rs$, its entanglement transfers from $\tilde b$ to $E$.  But general principles of entanglement say that this cannot happen if we just have local processes acting within the black hole, on $b$, and on $E$, we need something nonlocal.  Initially this is the alternative that seemed to me most likely.
In AdS/CFT, and holography more generally, locality in the bulk is emergent, not fundamental, so why shouldn't some subtle nonlocality be present even outside the black hole?

But actually one need a rather gross violation of EFT, an order one effect on the propagation.  One example would be a model where the mode $b$ produced by the Hawking process is simply reabsorbed into the black hole when it reaches the radius $2\rs$, and a new $b$, entangled with $E$, appears there.  Other models are being explored by Giddings~\cite{Giddings:2012gc}; one class would have additional information-emitting processes that offset the effect of the Hawking process.\footnote{We follow Giddings' term `Non-Violent Non-Locality.'}  However, it seems that one could still encounter a paradox.  For example, these processes would have to use different channels from the Hawking process, like higher partial waves, and one could surround the black hole with a mirror that reflects these back.  

Based on this kind of difficulty, and the large magnitude of the necessary effect, this idea seems unlikely to me.  (I believe that it is also disfavored by the argument based on chaos to be given later).  Still, of all the scenarios, this is the most interesting in one sense: any such model would likely have an order one effect on the dynamics of black holes at distances less than $O(2\rs)$, which would be observable.

\subsection{Fiery/braney/stringy drama}

So what if we give up No Drama?  If $b$ is not entangled with $\tilde b$, then we are not in the $a$-vacuum, and there is an excitation where the adiabatic principle would not allow one.  How energetic is this?  Basically, there is no upper limit due to the blue shifting, the earlier we meet the mode, the higher its energy, up to whatever cutoff we wish to impose on EFT.  And how many such photons are there?  Here there is a question.  Due to the greybody factors, the mismatch $2 S_b$ in strong subadditivity is of order $e^{-\ell}$.  So $\ell$ does not have to be very large before this is small enough that tiny effects remove the discrepancy.  So the minimal case is that we get an `$s$-wave firewall,' with excitations only in the low-$\ell$ modes.  

But there are still problems.  We can defeat the centrifugal barrier with a mining experiment, instead of waiting for the mode to come out we go down and look at it, and then we have the same paradox.  Mining experiments seem complicated, but have a long history in this subject.  This would lead to the conclusion that high energy excitations are present in all modes, and we burn up if we try to pass through.  This is more dramatic than the $s$-wave, but even that is a violation of the adiabatic principle.

So if drama at the horizon is the answer, how does the adiabatic principle get violated?  In particular, can we see some sign of this using stringy ingredients? Recall that with the original paradox, progress was made only after branes were brought into the story.  One scenario (and we're talking scenarios from here on out) is the fuzzball:  the singularity expands out into a braney shell at the horizon~\cite{Mathur:2009hf}.  We know of other places where this happens.  For example, if you break some of the symmetry of the ${\cal N}=4$ theory, such as adding masses for some of the QFT fields, this is a perturbation of the bulk fields that grows large as we go into the interior.  In some cases it leads to a singularity, which is resolved by expanding into branes (the enhan\'con, and Polchinski-Strassler).  But in those cases the singularity is timelike and the expansion is causal, while here the singularity is spacelike and must expand backwards in time.  But perhaps the expansion can be understood as a tunneling process~\cite{Kraus:2015zda,Mathur:2015nra}.  Just to be clear, one would be running into a shell of branes rather than a shell of high energy particles.

Silverstein and Dodelson~\cite{Silverstein:2014yza}  are looking at string scattering and production near the horizon, and have found interesting results. At the black hole horizon there are no large local invariants, but there are large nonlocal invariants, a large relative boost between the infalling matter and later infalling observers.  In quantum field theory, only large local invariants lead to breakdown, but strings have some funny properties due to their internal dynamics, when they are highly boosted they don't behave like point particles.    It is notable that string quantization in a black hole environment has never really been developed.  String perturbation theory works for the $S$-matrix and other on-shell questions, but when some particles fall through into the singularity there is no asymptotic state.  If we could do a nice-slice quantization of strings we could perhaps justify the adiabatic principle and rule out a large effect, but it is hard to quantize strings in anything but light-cone time, so there may be something interesting.

Finally, since we are thinking that spacetime is emergent, we might try the slogan that it is not that the firewall appears, but that the interior spacetime fails to emerge.  But to claim this we would need a better understanding of emergent spacetime.  Maybe this will come from recent developments in entanglement entropy.

Would firewalls change the result of the Hawking calculation?  One argument to the contrary is causality. Although the calculation as done assumes a smooth horizon, nothing that happens behind the horizon can affect anything that emerges outside it.  Another argument is the universal properties of high energy states in complex systems: the population of the outgoing mode $b$ should then be thermal on general grounds.

\subsection{Quantum drama}

Most attempts to avoid the firewall conclusion have introduced a different kind of drama.  Rather than a physical disturbance at the horizon, these change the rules of quantum mechanics for the infalling observer.  This is different from what Hawking proposed, which would affect observations by observers outside the black hole, the result of the swap test in particular.

As one example of quantum drama, suppose I give you 100 spin-$\frac12$ particles, and tell you that I have prepared them in such a way that if you measure $s_z$ for any of them you get $+\frac12$, and if you measure $s_n$ along a different axis $n$ you also get $+\frac12$.  Now, you know that I can't do this, if $s_z$ is definitely $+\frac12$ then along any other axis there will be some probability for $+\frac12$ and some for $-\frac12$.  But you choose half at random and measure $s_z$ for those, and $s_n$ for the others, and it comes out as I assert, $+\frac 12$ for all.  Now this is strange, much stranger than just burning up at the horizon.

But this is what you can get if you impose a future boundary condition.  The usual QM rule is that the probability of an observation $a$ (say $+\frac12$ along some chosen axis) is 
\be
{\rm Tr}(P_a\rho) \,,
\ee
but with a specified final state $b$ it becomes
\be
{\rm Tr}(P_a P_b P_a\rho) \,.
\ee
Now, for $\rho$ a projection on the $z$-axis, and $P_b$ a projection on the $n$-axis, you get the stated result.  In fact, you can even measure $s_z$ first and then $s_n$, and still get the same effect:
\be
{\rm Tr}(P_z P_n P_b P_n P_z \rho) = {\rm Tr}(P_z P_b P_z\rho) = {\rm Tr}(P_n P_b P_n\rho) = {\rm Tr}(P_b\rho)  \,,
\ee
but not if you measure $s_n$ and then $s_z$!

In the black hole context, this idea was suggested some time ago by Horowitz and Maldacena~\cite{Horowitz:2003he}.  The logic was that we expect a definite state at an initial singularity, so perhaps this should hold at the black hole final singularity as well.  This looks like it might cure the information problem, by projecting to a definite state on the interior of the nice slice.  And it avoids the firewall, because it allows both the $\tilde bb$ and $bE$ entanglements, by analog to the above argument with spins.  Of course, one has to wonder how causality works for the infalling observer, and closer inspection points to difficulties, including acausality leaking out to the black hole exterior~\cite{Bousso:2013uka,Lloyd:2013bza}.

Another possibility that I would classify as quantum drama is that one cannot actually perform the AMPS experiment because it takes much too long~\cite{Harlow:2013tf}.  Counterarguments have also been given, based on thought experiments in AdS~\cite{AMPSS} and on the use of precomputation~\cite{OU}.  But beyond this, it is one thing to find an operational interpretation, and another to base a theory on it.  What does it mean that a state can violate a basic quantum mechanical identity as long as no observer can detect it?  This would be some modification of quantum theory that falls under the category of quantum drama.

I will touch on other ideas, state dependence and ER=EPR,  in the following sections.

\section{Bulk reconstruction}

\subsection{Precursors}

What about AdS/CFT?  This was supposed to be a complete theory of quantum gravity, able to answer all questions.  Can't this tell us whether there is a firewall?  

One might think that a firewall would look rather dramatic in the CFT.  But it would sit near the horizon, where $g_{tt} = 0$ and there is a big redshift: energies will look much smaller in the CFT.  Further, we expect all states to look basically thermal in any case, by the considerations in \S5.3.  So we need to understand better how the bulk physics is encoded in the CFT.

Note that to answer the question about information loss, we did not even need to solve the CFT, just to know that it existed.  But with firewalls, even if we could solve the CFT, it is not clear what question to ask.  We have a sharp dictionary only for modes near the boundary.  This tells us the black hole S-matrix, but it does not immediately tell us what happens in the interior.  So I will review what is known about reconstructing the bulk from the CFT even outside the black hole.  There has been interesting recent progress here, based on the idea of quantum error correction, which I will discuss at the end. 

Suppose I prepare the system in some state and give you the CFT description.  What measurements would you do on the state, to determine the presence of excitations in the bulk?  The GKPW dictionary~\cite{GKP,W} relates local bulk operators to fields near the boundary.  So if our state looks like the vacuum plus local gauge-invariant excitations, then we know that we have bulk excitations near the boundary.  But what if we have excitations further in? 

A typical CFT operator is something like Tr$(F_{\mu\nu}(x) F^{\mu\nu}(x))$.  After it acts, the two excitations in the CFT begin to move away at the speed of light in independent directions.  Because of the strong interactions, they will also fragment in to more quanta.  In the bulk picture, the excitations created by ${\cal O}_i$ move into the boundary, also at the speed of light if we are talking about an operator dual to a supergravity field.  So to see an excitation at some radial coordinate $r$, we would look for a non-vacuum correlation in the fields over some distance which is roughly reciprocal to $r$ (as long as $r \gg l_{\rm AdS}$, at smaller $r$ it is not so simple).

To give an explicit construction~\cite{Hamilton:2006az}  for these {\it precursors} \cite{Polchinski:1999yd}\ (so-named because the time-reversed case was considered first) or {\it bulk fields}, begin with the bulk equation of motion, e.g. Klein-Gordon equation
\be
(\partial^2 - m_i^2) \phi_i = 0 \,.  \label{kge}
\ee
Integrating this, we can express $\phi_i$ at interior points in terms of its boundary values, and then in terms of CFT operators via the dictionary~(\ref{extrap}):
\be
\phi(r, t, \theta) = \int dt'd\theta'\, K(r, t, \theta';  t, \theta') {\cal O}_i(t',\theta') \,, \label{precur}
\ee
where the $K$ are known as smearing functions.  The RHS here consists of local operators at various times, but by using the CFT equations of motion we can express them in terms of nonlocal operators at some reference time, say $t$.  The smearing operators we can give explicitly, but the single-time nonlocal form requires that we can solve the CFT, so this is implicit.  It is sometimes useful to consider a free-field model of the CFT, although there is not expected to be true local physics in the bulk in this case.

Integrating to a timelike boundary is not a standard causal problem.  We can write the smearing functions by using spherical symmetry to reduce to a 1+1 dimensional problem, for which space and time can be reversed.  The resulting smearing function has support on all boundary points spacelike with respect to the bulk point $p = (r, t, \theta)$.  However, due to the non-Cauchy nature of the problem, $K$ is not unique.  The operators ${\cal O}_i(t',\theta')$ satisfy dynamical relations (periodicity in AdS time, and others) such that different $K$ give the same operator.  Using this, the support of $K$ can be changed. 

The free field equation~(\ref{kge}) gives the precursor to leading order in $1/N$.  It can be systematically improved by including the bulk interactions, leading to the form
\bea
\phi_i(r, t, \theta) &=& \int dt'd\theta'\, K(r, t, \theta;  t', \theta') {\cal O}_i(t',\theta') \nonumber\\
&&\quad + \frac{1}{N} \int dt'd\theta' dt''d\theta''\, K_{ijk}(r, t, \theta;  t', \theta', t'', \theta'') {\cal O}_j(t',\theta'){\cal O}_k(t'',\theta'') + \ldots \,.
\label{preser}
\eea

Note that this construction is requiring us to solve the bulk dynamics explicitly; the dual gauge theory is not doing all the work.  As we get to more complicated situations, this means that if we can't solve the bulk dynamics, we can't identify the precursor, at least by this process.  It has been suggested that spacelike commutativity of the bulk fields (at least to the extent allowed by bulk gauge invariances) might identify them, but it seems that one needs dynamical information as well.   As we will discuss below, entanglement might make possible a more intrinsic construction of the precursor.

The field equations are being integrated in a given background such as AdS.  However, the result~(\ref{preser}) is an operator equation: a change in the background means a change in the expectation values of ${\cal O}_i$, and so the nonlinear terms should restore background independence.  As written this would seem to be true only within the radius of convergence of the sum.  However, we expect that it can then be extended further, in the spirit of analytic continuation, until a natural boundary is reached.  The only natural boundary of which we are aware is the horizon of a black hole.

The problem with a black hole horizon is that the integration of the field equation hits the singularity, so cannot be related to boundary operators.  Or we can take advantage of the ambiguity of $K$ to integrate back in time and then to the boundary.  But this only works for very young black holes, less than the scrambling time, else the blue shift leads to trans-Planckian dynamics.

When the bulk state approaches the horizon, the precursor construction breaks down and we must understand the quantum theory more completely in order to probe the black hole interior.  We discuss here two directions, state-dependent and ER=EPR.

\subsection{State dependence}

Papadodimas and Raju have proposed a `state-dependent' construction of operators in the black hole interior, which moreover show no firewall~\cite{Papadodimas:2012aq}.  The first step is to note that since typical states\footnote{Haar-random states in some ensemble of limited energy.} look thermal,   
the distribution of the Hawking mode will be thermal.  But also for infalling vacuum, the Hawking calculation gives a thermal distribution.  So it seems consistent to postulate that typical states are infalling vacuum, as effective field theory gives.  One can then use the entanglement pattern $b\tilde b$ to identify the states of $\tilde b$ and construct the interior operators.  But there is a puzzle: typical states are vacuum, but every vacuum state is associated with a Fock space of $\tilde b$ excitations.  How are we to fit all of these into the Hilbert space?

To do this we have to squash the Hilbert space in a nonlinear way.  In normal quantum mechanics, physically distinct states (e.g. infalling vacuum, infalling excitation) are orthogonal.  In order to fit the interior Hilbert space, without firewalls, the PR proposal requires that physically orthogonal states not always be orthogonal, and in fact they are often quite close to parallel.  This is made precise in~\cite{Marolf:2015dia}.

This nonlinearity is another form of quantum drama.  The nonlinearity of observables on Hilbert space has come to be called `state-dependence,' but it is much different from more normal forms of background-dependence such as the precursors. That construction is nonlinear in $\cal O$, but the product of linear operators is still a linear operator.  It may be that the proposed modification of quantum mechanics is a feature not a bug, a necessary property of quantum gravity.  But any modification of quantum mechanics requires a great deal of infrastructure to be rethought. 

\subsection{ER=EPR}

Thus far we have focused on a single black hole formed in decay.  Let us consider now a pair of black holes in equilibriam.  The combined state is taken to be pure, while the two holes are highly entangled with one another,
\be
| \Psi \rangle = \sum_n e^{-\beta E_n/2} |n,n\rangle  \,.
\ee
Geometrically, this describes the full Kruskal metric, all four quadrants~(\ref{quad}) connected through an Einstein-Rosen bridge.  A typical state with such a geometry will be highly entangled, a situation summarized at ER $\rightarrow$ EPR: the two-sided geometry is highly entangled.

The question is, does this go in both directions, EPR $\rightarrow$ ER?  Does high entanglement imply the two-sided connected geometry~\cite{Maldacena:2013xja}, giving the two-sided interence ER = ERP?  This may seem natural.  It so, there is no firewall: the geometry connects the two sides smoothly.  But this is in a way radical.  

Consider starting with a single black hole.  Let half decay, forming an entangled pair.  Using quantum computation, manipulating only the bits on one side, 
bring the pair to the typical coupled system.  Geometrically, this is the two-sided black hole.  So, starting from two physically separated systems which are connected in some complicated way, we can by acting on either side, bring them to a state connected by an ER bridge, EPR $\rightarrow$ ER, so the inference is two-sided.  

The smooth bridge implies no firewall: the infalling observer passes smoothly into future quadrant 2.  But this is very strange.  Quadrant 2 receives messages from both quadrant 1 and quadrant 3, the former being the half-decayed black hole and the latter being the Hawking radiation, recaptured and quantum-computed to give a second copy.  An observer jumping in from region 1 and receive messages from region 3, even though these seem to be completely causally disconnected.

\section{Moving forward}

Before concluding, I want to mention to very interesting recent ideas, which are not directly aimed at the information problem but which are certainly connected with that circle of ideas.  The first is a new approach to the precursor construction, and the second it the role of chaos. 

\subsection{Quantum information}

In \S8.1 we discussed the difficulties in reconstructing the bulk physics in the CFT, especially as we approach the horizon.  We noted that the bulk construction is a non-standard Cauchy problem.  Refs.~\cite{Almheiri:2014lwa,Pastawski:2015qua} recast this as a question in quantum computation, quantum secret sharing in particular, where it may be more natural.
Considering a boundary $S^1$, the precursor construction allows one to construct the field in the center in terms of CFT operators localized in any arc of angle greater than $\pi$.  For example, dividing the boundary into thirds $A,B,C$, one can reconstruct the state in the center from any two regions; any one can be deleted and the information is not lost.

This is a new way to look at precursors, and it may be more powerful.
In the standard precursor construction, given a region of the CFT one can reconstruct the so-called causal wedge of the bulk.  Refs.~\cite{Almheiri:2014lwa,Pastawski:2015qua,Jafferis:2015del,Dong:2016eik,Harlow:2016vwg} suggest that quantum error correction allows one to reach a larger area, the entanglement wedge.  If this works, it seems very deep.  Since AdS/CFT we have developed the technology of the precursor construct, but this seems like a really conceptual change.  But it remains to be seen whether we can use it to see behind the horizon, and what it will say about the dynamics there.

\subsection{Chaos}

Thermal behavior and chaos are intimately connected. In thermalizing systems, the ergodic mixing of the phase space arises from the exponential divergence of nearby trajectories.  It has been recognized for more than four decades that black holes have thermodynamic properties. However, only very recently has the connection with chaos been made~\cite{Shenker:2013pqa,Shenker:2014cwa,kitaev,Maldacena:2015waa}.   

Chaos is another aspect of entanglement that has recently come to the fore.  It can be seen as another argument for the firewall paradox~\cite{Almheiri:2013hfa,Shenker:2013pqa}.    It leads to interesting questions, and connections, between quantum behavior and the black hole dynamics~\cite{Shenker:2013pqa,Shenker:2014cwa,kitaev,Maldacena:2015waa}.  As a quantum field theorist, chaos is notable because the study of its dynamics requires the use of the four-point function out of time ordering, which is unfamiliar to most of us.  And, a recent two-dimensional chaotic model~\cite{kitaev,Sachdev:1992fk,Polchinski:2016xgd} provised a very interesting model of holography.
  
I became interested in this~\cite{Polchinski:2015cea} in connection with an old puzzle about 't Hooft~\cite{Dray:1984ha,'tHooft:1996tq}.  In particular, he introduces the idea of a calculable black hole S-matrix, which moreover has stringy properties.   It is based on the calculation of the scattering of right- and left-moving near horizon modes.  But there should be no right-moving modes, by the adiabatic principle, so this S-matrix should not exist.  In~\cite{Polchinski:2015cea} I argue that this S-matrix can be given a more limited interpretation in terms of chaos, using the out-of-time order property.  Moreover, we conclude that chaos again leads to the necessity of the firewall.

Finally, for your interest, I mention the work~\cite{Brown:2015bva}, which goes beyond chaos to complexity.  I have no intuition for this as year.
 
  \subsection{Conclusions}

Let us first review.
I have talked about the three classic alternatives.  One can divide these further, e.g. stable versus long-lived remnants, but all ideas that come up seem to fit into one of these three general frameworks.  The situation today is not so clear.  

I count at least {\it ten} possibilities.  Final state QM, ER=EPR, state-dependence, limits on quantum computation, strong complementarity, and the classic Page- and energy-violation represent six distinct ways that we might modify quantum mechanics so as to avoid the firewall.  Are any consistent?  The other four are modifications of the horizon:  fuzzballs, fire, strings, and NVNL.  Some of these (such as NVNL) are conceived as affecting the geometry out to the horizon scale, while others (such as the firewall) appear only near the horizon.  I have always tended to be an agnostic, considering all possible solutions as long at they are consistent.  

So how are we to make progress? On the one hand, what we are doing in this lecture, identifying the general possibilities and constraints, is important.  But likely we also need new ideas from unexpected directions, as with AdS/CFT.  Students often ask, what should I calculate?  With black hole information it can be difficult, it is very conceptual.  

With AdS/CFT, there was a wealth of calculation of brane dynamics, which presaged first the black hole state counting and then the duality.
Today, there is a wealth of calculation going on in the area of quantum information and entanglement.  
Clearly this is very rich.  It may have many applications, but I urge you to remember the big question, `what is quantum gravity.'  Are we close to an answer, with all of our new quantum toys?  Or are we still far away, with our limitation to the holographic boundary?  Do we perhaps still miss ideas that are as large as those that we have understood?


\section*{Acknowledgments}

I am grateful for my many collaborators, students, and colleagues over the years.  I thank David Gross and
Eliezer Rabinovici for inviting me to the 2014-15 Jerusalem Winter School, and my collaborators
  Pedro Vieira, Oliver DeWolfe, and Tom DeGrand on the 2015  TASI school.  This work was supported by NSF Grant PHY11-25915 (academic year) and PHY13-16748 (summer).



\begin{thebibliography}{99}
\itemsep = 3pt

\bibitem{Wilson:1993dy} 
  K.~G.~Wilson,
  ``The renormalization group and critical phenomena,''
  Rev.\ Mod.\ Phys.\  {\bf 55}, 583 (1983).
  
\bibitem{Witten:1985cc} 
  E.~Witten,
  ``Noncommutative Geometry and String Field Theory,''
  Nucl.\ Phys.\ B {\bf 268}, 253 (1986).
  
\bibitem{Banks:1996vh} 
  T.~Banks, W.~Fischler, S.~H.~Shenker and L.~Susskind,
  ``M theory as a matrix model: A Conjecture,''
  Phys.\ Rev.\ D {\bf 55}, 5112 (1997)
  [hep-th/9610043].

\bibitem{Maldacena:1997re}
  J.~M.~Maldacena,
  ``The large N limit of superconformal field theories and supergravity,''
  Adv.\ Theor.\ Math.\ Phys.\  {\bf 2}, 231 (1998)
  [Int.\ J.\ Theor.\ Phys.\  {\bf 38}, 1113 (1999)]
  [arXiv:hep-th/9711200].

\bibitem{GKP}
  S.~S.~Gubser, I.~R.~Klebanov and A.~M.~Polyakov,
  ``Gauge theory correlators from non-critical string theory,''
  Phys.\ Lett.\  B {\bf 428}, 105 (1998)
  [arXiv:hep-th/9802109].

\bibitem{W}
  E.~Witten,
  ``Anti-de Sitter space and holography,''
  Adv.\ Theor.\ Math.\ Phys.\  {\bf 2}, 253 (1998)
  [arXiv:hep-th/9802150].
  
\bibitem{'tHooft:1993gx}
  G.~'t Hooft,
  ``Dimensional reduction in quantum gravity,''
  arXiv:gr-qc/9310026.

\bibitem{Susskind:1994vu}
  L.~Susskind,
  ``The World As A Hologram,''
  J.\ Math.\ Phys.\  {\bf 36}, 6377 (1995)
  [arXiv:hep-th/9409089].
  
\bibitem{Strominger:1996sh} 
  A.~Strominger and C.~Vafa,
  ``Microscopic origin of the Bekenstein-Hawking entropy,''
  Phys.\ Lett.\ B {\bf 379}, 99 (1996)
  [hep-th/9601029].
  
\bibitem{Hawking:1976ra} 
  S.~W.~Hawking,
  ``Breakdown of Predictability in Gravitational Collapse,''
  Phys.\ Rev.\ D {\bf 14}, 2460 (1976).

\bibitem{Almheiri:2012rt} 
  A.~Almheiri, D.~Marolf, J.~Polchinski and J.~Sully,
  ``Black Holes: Complementarity or Firewalls?,''
  JHEP {\bf 1302}, 062 (2013)
  [arXiv:1207.3123 [hep-th]].

\bibitem{Ryu:2006bv} 
  S.~Ryu and T.~Takayanagi,
  ``Holographic derivation of entanglement entropy from AdS/CFT,''
  Phys.\ Rev.\ Lett.\  {\bf 96}, 181602 (2006)
  [hep-th/0603001].
  
 \bibitem{Harlow:2014yka} 
  D.~Harlow,
  ``Jerusalem Lectures on Black Holes and Quantum Information,''
  arXiv:1409.1231 [hep-th].

\bibitem{Page:1993up} 
  D.~N.~Page,
  ``Black hole information,''
  hep-th/9305040.
  
\bibitem{Mathur:2009hf}
  S.~D.~Mathur,
  ``The Information paradox: A Pedagogical introduction,''
  Class.\ Quant.\ Grav.\  {\bf 26}, 224001 (2009)
  [arXiv:0909.1038 [hep-th]];

\bibitem{Bekenstein:1973ur} 
  J.~D.~Bekenstein,
  ``Black holes and entropy,''
  Phys.\ Rev.\ D {\bf 7}, 2333 (1973).
  
\bibitem{Hawking:1974sw} 
  S.~W.~Hawking,
  ``Particle Creation by Black Holes,''
  Commun.\ Math.\ Phys.\  {\bf 43}, 199 (1975)
  [Erratum-ibid.\  {\bf 46}, 206 (1976)].

\bibitem{Jacobson:2003vx} 
  T.~Jacobson,
  ``Introduction to quantum fields in curved space-time and the Hawking effect,''
  gr-qc/0308048.
  
\bibitem{Mukhanov:2007zz} 
  V.~Mukhanov and S.~Winitzki,
  ``Introduction to quantum effects in gravity,''
  Cambridge, UK: Cambridge Univ. Pr. (2007) 273 p
    
\bibitem{Giddings:1992ff} 
  S.~B.~Giddings and W.~M.~Nelson,
  ``Quantum emission from two-dimensional black holes,''
  Phys.\ Rev.\ D {\bf 46}, 2486 (1992)
  [hep-th/9204072].
  
  \bibitem{Page:1993wv} 
  D.~N.~Page,
  ``Information in black hole radiation,''
  Phys.\ Rev.\ Lett.\  {\bf 71}, 3743 (1993)
  [hep-th/9306083].
  
  \bibitem{Ellis:1983jz} 
  J.~R.~Ellis, J.~S.~Hagelin, D.~V.~Nanopoulos and M.~Srednicki,
  ``Search for Violations of Quantum Mechanics,''
  Nucl.\ Phys.\ B {\bf 241}, 381 (1984).
  
\bibitem{Banks:1983by} 
  T.~Banks, L.~Susskind and M.~E.~Peskin,
  ``Difficulties for the Evolution of Pure States Into Mixed States,''
  Nucl.\ Phys.\ B {\bf 244}, 125 (1984).
  
  
\bibitem{Strominger:1994ey} 
  A.~Strominger,
  ``Unitary rules for black hole evaporation,''
  hep-th/9410187. 
    
\bibitem{Oppenheim:2009nb} 
  J.~Oppenheim and B.~Reznik,
  ``Fundamental destruction of information and conservation laws,''
  arXiv:0902.2361 [hep-th].
 
\bibitem{Unruh:2012vd} 
  W.~G.~Unruh,
  ``Decoherence without Dissipation,''
  Trans.\ Roy.\ Soc.\ Lond.\  {\bf 370}, 4454 (2012)
  [arXiv:1205.6750 [quant-ph]].
  
 \bibitem{Polchinski:2010hw} 
  J.~Polchinski,
  ``Introduction to Gauge/Gravity Duality, TASI 2010,"
   arXiv:1010.6134 [hep-th].
  
  
\bibitem{Maldacena:2001kr} 
  J.~M.~Maldacena,
  ``Eternal black holes in anti-de Sitter,''
  JHEP {\bf 0304}, 021 (2003)
  doi:10.1088/1126-6708/2003/04/021
  [hep-th/0106112].
  
\bibitem{Horowitz:1999uv} 
  G.~T.~Horowitz,
  ``Comments on black holes in string theory,''
  Class.\ Quant.\ Grav.\  {\bf 17}, 1107 (2000)
  [hep-th/9910082].
  
\bibitem{Witten:1998zw} 
  E.~Witten,
  ``Anti-de Sitter space, thermal phase transition, and confinement in gauge theories,''
  Adv.\ Theor.\ Math.\ Phys.\  {\bf 2}, 505 (1998)
  [hep-th/9803131].
  
\bibitem{ETH}
J. M. Deutsch, Phys. Rev. A 43, 2046 (1991);

M. Srednicki, Phys. Rev. E 50, 888 (1994) [arXiv:cond-mat/9406056]; J. Phys. A 29, L75 (1996)
[arXiv:cond-mat/9406056]; J. Phys. A 32, 1163 (1999) [arXiv:cond-mat/9809360].

\bibitem{Hayden:2007cs} 
  P.~Hayden and J.~Preskill,
  ``Black holes as mirrors: Quantum information in random subsystems,''
  JHEP {\bf 0709}, 120 (2007)
  [arXiv:0708.4025 [hep-th]].
  
  
\bibitem{rislam}
R. Islam, et al., in preparation.


\bibitem{Susskind:1993if}
  L.~Susskind, L.~Thorlacius and J.~Uglum,
  ``The Stretched horizon and black hole complementarity,''
  Phys.\ Rev.\ D {\bf 48}, 3743 (1993)
  [hep-th/9306069].

 \bibitem{Stephens:1993an}
  C.~R.~Stephens, G.~'t Hooft and B.~F.~Whiting,
  ``Black hole evaporation without information loss,''
  Class.\ Quant.\ Grav.\  {\bf 11}, 621 (1994)
  [gr-qc/9310006].

\bibitem{Susskind:1993mu}
  L.~Susskind and L.~Thorlacius,
  ``Gedanken experiments involving black holes,''
  Phys.\ Rev.\ D {\bf 49}, 966 (1994)
  [hep-th/9308100].

\bibitem{Preskill}
J.~Preskill, quoted in Ref.~\cite{Susskind:1993if}



\bibitem{Sekino:2008he}
  Y.~Sekino and L.~Susskind,
  ``Fast Scramblers,''
  JHEP {\bf 0810}, 065 (2008)
  [arXiv:0808.2096 [hep-th]].
  

\bibitem{Giddings:2011ks}
  S.~B.~Giddings,
  ``Models for unitary black hole disintegration,''
  Phys.\ Rev.\ D {\bf 85}, 044038 (2012)
  [arXiv:1108.2015 [hep-th]];
    S.~B.~Giddings and Y.~Shi,
  ``Quantum information transfer and models for black hole mechanics,''
  arXiv:1205.4732 [hep-th].

\bibitem{Giddings:2012gc} 
  S.~B.~Giddings,
  ``Nonviolent nonlocality,''
  Phys.\ Rev.\ D {\bf 88}, 064023 (2013)
  [arXiv:1211.7070 [hep-th]].

 
\bibitem{Almheiri:2013hfa} 
  A.~Almheiri, D.~Marolf, J.~Polchinski, D.~Stanford and J.~Sully,
  ``An Apologia for Firewalls,''
  JHEP {\bf 1309}, 018 (2013)
  [arXiv:1304.6483 [hep-th]].
    
\bibitem{Bousso:2013wia} 
  R.~Bousso,
  ``Firewalls from double purity,''
  Phys.\ Rev.\ D {\bf 88}, no. 8, 084035 (2013)
  [arXiv:1308.2665 [hep-th]].
  
\bibitem{Marolf:2013dba} 
  D.~Marolf and J.~Polchinski,
  ``Gauge/Gravity Duality and the Black Hole Interior,''
  Phys.\ Rev.\ Lett.\  {\bf 111}, 171301 (2013)
  doi:10.1103/PhysRevLett.111.171301
  [arXiv:1307.4706 [hep-th]].

\bibitem{Unruh:1995gn} 
  W.~G.~Unruh and R.~M.~Wald,
  ``On evolution laws taking pure states to mixed states in quantum field theory,''
  Phys.\ Rev.\ D {\bf 52}, 2176 (1995)
  [hep-th/9503024].
  
\bibitem{Chen:2014jwq} 
  P.~Chen, Y.~C.~Ong and D.~h.~Yeom,
  ``Black Hole Remnants and the Information Loss Paradox,''
  arXiv:1412.8366 [gr-qc].
 
\bibitem{Kraus:2015zda} 
  P.~Kraus and S.~D.~Mathur,
  ``Nature abhors a horizon,''
  arXiv:1505.05078 [hep-th].
  
\bibitem{Mathur:2015nra} 
  S.~D.~Mathur,
  ``A model with no firewall,''
  arXiv:1506.04342 [hep-th].
  
\bibitem{Silverstein:2014yza} 
  E.~Silverstein,
  ``Backdraft: String Creation in an Old Schwarzschild Black Hole,''
  arXiv:1402.1486 [hep-th].
  
M.~Dodelson and E.~Silverstein,
  ``String-theoretic breakdown of effective field theory near black hole horizons,''
  arXiv:1504.05536 [hep-th].

 M.~Dodelson and E.~Silverstein,
  ``Longitudinal nonlocality in the string S-matrix,''
  arXiv:1504.05537 [hep-th].
  
\bibitem{Horowitz:2003he} 
  G.~T.~Horowitz and J.~M.~Maldacena,
  ``The Black hole final state,''
  JHEP {\bf 0402}, 008 (2004)
  [hep-th/0310281].
  
\bibitem{Bousso:2013uka} 
  R.~Bousso and D.~Stanford,
  ``Measurements without Probabilities in the Final State Proposal,''
  Phys.\ Rev.\ D {\bf 89}, no. 4, 044038 (2014)
  [arXiv:1310.7457 [hep-th]].
  
\bibitem{Lloyd:2013bza} 
  S.~Lloyd and J.~Preskill,
  ``Unitarity of black hole evaporation in final-state projection models,''
  JHEP {\bf 1408}, 126 (2014)
  [arXiv:1308.4209 [hep-th]].
  
\bibitem{Harlow:2013tf}
  D.~Harlow and P.~Hayden,
  ``Quantum Computation vs. Firewalls,''
  arXiv:1301.4504 [hep-th].
  
 \bibitem{AMPSS}
A.~Almheiri, D.~Marolf, J.~Polchinski, D.~Stanford and J.~Sully,
  ``An Apologia for Firewalls,''
  JHEP {\bf 1309}, 018 (2013)
  [arXiv:1304.6483 [hep-th]];

\bibitem{OU}
J. Oppenheim and W. G. Unruh, 
``Firewalls and 
at mirrors: An alternative to the
AMPS experiment which evades the Harlow-Hayden obstacle,"
 JHEP 1403 (2014) 120,
[arXiv:1401.1523].

\bibitem{Hamilton:2006az} 
  A.~Hamilton, D.~N.~Kabat, G.~Lifschytz and D.~A.~Lowe,
  ``Holographic representation of local bulk operators,''
  Phys.\ Rev.\ D {\bf 74}, 066009 (2006)
  [hep-th/0606141].

\bibitem{Polchinski:1999yd} 
  J.~Polchinski, L.~Susskind and N.~Toumbas,
  ``Negative energy, superluminosity and holography,''
  Phys.\ Rev.\ D {\bf 60}, 084006 (1999)
  [hep-th/9903228].

\bibitem{Papadodimas:2012aq} 
  K.~Papadodimas and S.~Raju,
  ``An Infalling Observer in AdS/CFT,''
  JHEP {\bf 1310}, 212 (2013)
  [arXiv:1211.6767 [hep-th]].
  
\bibitem{Marolf:2015dia} 
  D.~Marolf and J.~Polchinski,
  ``Violations of the Born rule in cool state-dependent horizons,''
  JHEP {\bf 1601}, 008 (2016)
  doi:10.1007/JHEP01(2016)008
  [arXiv:1506.01337 [hep-th]].
  
\bibitem{Maldacena:2013xja} 
  J.~Maldacena and L.~Susskind,
  ``Cool horizons for entangled black holes,''
  Fortsch.\ Phys.\  {\bf 61}, 781 (2013)
   [arXiv:1306.0533 [hep-th]].

\bibitem{Almheiri:2014lwa} 
  A.~Almheiri, X.~Dong and D.~Harlow,
  ``Bulk Locality and Quantum Error Correction in AdS/CFT,''
  JHEP {\bf 1504}, 163 (2015)
  [arXiv:1411.7041 [hep-th]].
  
\bibitem{Pastawski:2015qua} 
  F.~Pastawski, B.~Yoshida, D.~Harlow and J.~Preskill,
  ``Holographic quantum error-correcting codes: Toy models for the bulk/boundary correspondence,''
  arXiv:1503.06237 [hep-th].
  
\bibitem{Jafferis:2015del} 
  D.~L.~Jafferis, A.~Lewkowycz, J.~Maldacena and S.~J.~Suh,
  ``Relative entropy equals bulk relative entropy,''
  JHEP {\bf 1606}, 004 (2016)
  [arXiv:1512.06431 [hep-th]].
  
\bibitem{Dong:2016eik} 
  X.~Dong, D.~Harlow and A.~C.~Wall,
  ``Reconstruction of Bulk Operators within the Entanglement Wedge in Gauge-Gravity Duality,''
  Phys.\ Rev.\ Lett.\  {\bf 117}, no. 2, 021601 (2016)
  [arXiv:1601.05416 [hep-th]]. 
  
\bibitem{Harlow:2016vwg} 
  D.~Harlow,
  ``The Ryu-Takayanagi Formula from Quantum Error Correction,''
  arXiv:1607.03901 [hep-th].

\bibitem{Shenker:2013pqa} 
  S.~H.~Shenker and D.~Stanford,
  ``Black holes and the butterfly effect,''
  JHEP {\bf 1403}, 067 (2014)
  [arXiv:1306.0622 [hep-th]].
   
\bibitem{Shenker:2014cwa} 
  S.~H.~Shenker and D.~Stanford,
  ``Stringy effects in scrambling,''
  JHEP {\bf 1505}, 132 (2015)
   [arXiv:1412.6087 [hep-th]].
  
\bibitem{kitaev}
A.~Kitaev,
``Hidden correlations in the Hawking radiation and thermal noise,'' Breakthrough Prize Fundamental Physics Symposium 11/10/2014, KITP seminar 2/12/2015; 	``A simple model of quantum holography (part 1),'' KITP seminar 4/7/2015.
  

\bibitem{Maldacena:2015waa} 
  J.~Maldacena, S.~H.~Shenker and D.~Stanford,
  ``A bound on chaos,''
  arXiv:1503.01409 [hep-th].
  
\bibitem{Sachdev:1992fk} 
  S.~Sachdev and J.~w.~Ye,
  ``Gapless spin fluid ground state in a random, quantum Heisenberg magnet,''
  Phys.\ Rev.\ Lett.\  {\bf 70}, 3339 (1993)
  doi:10.1103/PhysRevLett.70.3339
  [cond-mat/9212030].

\bibitem{Polchinski:2016xgd} 
  J.~Polchinski and V.~Rosenhaus,
  ``The Spectrum in the Sachdev-Ye-Kitaev Model,''
  JHEP {\bf 1604}, 001 (2016)
    [arXiv:1601.06768 [hep-th]]. 
   
\bibitem{Polchinski:2015cea} 
  J.~Polchinski,
  ``Chaos in the black hole S-matrix,''
  arXiv:1505.08108 [hep-th].

\bibitem{Dray:1984ha} 
  T.~Dray and G.~'t Hooft,
  ``The Gravitational Shock Wave of a Massless Particle,''
  Nucl.\ Phys.\ B {\bf 253}, 173 (1985).
  
\bibitem{'tHooft:1996tq} 
  G.~'t Hooft,
  ``The Scattering matrix approach for the quantum black hole: An Overview,''
  Int.\ J.\ Mod.\ Phys.\ A {\bf 11}, 4623 (1996)
  [gr-qc/9607022].

\bibitem{Brown:2015bva} 
  A.~R.~Brown, D.~A.~Roberts, L.~Susskind, B.~Swingle and Y.~Zhao,
  ``Holographic Complexity Equals Bulk Action?,''
  Phys.\ Rev.\ Lett.\  {\bf 116}, no. 19, 191301 (2016)
  [arXiv:1509.07876 [hep-th]].
  
 \end{thebibliography}
\end{document}